\documentclass[traditabstract]{aa}
\usepackage{psfrag,graphicx}
\usepackage{txfonts}
\usepackage{natbib}
\usepackage{ulem}
\usepackage{color}

\begin{document}
\bibliographystyle{aa}

\title{The implications of dust for high-redshift protogalaxies and the formation of binary disks}
\titlerunning{The chemistry and dynamics during the formation of high-redshift protogalaxies}
%   {Lyman alpha emission}

\author{M.~A.~Latif \inst{1}
\and
D.~R.~G.~Schleicher \inst{2}
\and
M.~Spaans \inst{1}
} 
%    \newauthor % starts a new line in the
\institute{Kapteyn Astronomical Institute, University of Groningen, P.O.~Box 800, 9700 AV Groningen, The Netherlands 
\and
% Institut fur Astrophysik, Georg-August-Universitat, Friedrich-Hund-Platz 37077 Gottingen 
Institut f\"ur Astrophysik, Georg-August-Universit\"at, Friedrich-Hund-Platz 1, 37077 G\"ottingen, Germany 
 }

\authorrunning{Latif et al.}

\date{today}

% \pagerange{\pageref{firstpage}--\pageref{lastpage}} \pubyear{2009}

% \def\LaTeX{L\kern-.36em\raise.3ex\hbox{a}\kern-.15em
%     T\kern-.1667em\lower.7ex\hbox{E}\kern-.125emX}

% \newtheorem{theorem}{Theorem}[section]

% \bibliographystyle{aa}

% \label{firstpage}
\newcommand{\HId}{$\mathrm{H}$}
\newcommand{\HIId}{$\mathrm{H}^+$}
\newcommand{\HzIId}{$\mathrm{H}_2^+$}
\newcommand{\HzId}{$\mathrm{H}_2$}
\newcommand{\HMd}{$\mathrm{H}^-$}
\newcommand{\HeMd}{$\mathrm{He}^-$}
\newcommand{\HeHIId}{$\mathrm{HeH}^+$}
\newcommand{\HeId}{$\mathrm{He}$}
\newcommand{\HeIId}{$\mathrm{He}^+$}
\newcommand{\HeIIId}{$\mathrm{He}^{++}$}
\newcommand{\DId}{$\mathrm{D}$}
\newcommand{\DIId}{$\mathrm{D}^+$}
\newcommand{\HDIId}{$\mathrm{HD}^+$}
\newcommand{\HDId}{$\mathrm{HD}$}
\newcommand{\DMd}{$\mathrm{D}^-$}
\newcommand{\ed}{$\mathrm{e}^-$}

\abstract{Numerical simulations suggest that the first galaxies are formed in protogalactic halos with virial temperatures $\rm \geq 10^{4}$ K. It is likely that such halos are polluted with trace amounts of metals produced by the first generation of stars. The presence of dust can significantly change the chemistry and dynamics of early galaxies. In this article, we aim to assess the role of dust on the thermal and dynamical evolution of the first galaxies in the presence of a background UV flux, and its implications for the observability of Lyman alpha emitters and sub-mm sources. We have performed high resolution cosmological simulations using the adaptive mesh refinement code FLASH to accomplish this goal. We have developed a chemical network appropriate for these conditions and coupled it with the FLASH code. The main ingredients of our chemical model include the formation of molecules (both in the gas phase and on dust grains), a multi-level treatment of atomic hydrogen, line trapping of Lyman alpha photons and, photoionization and photodissociation processes in a UV background. We found that the formation of molecules ($\rm H_{2}$ and HD) is significantly enhanced in the presence of dust grains as compared to only gas phase reactions by up to two orders of magnitude. The presence of dust may thus establish a molecular ISM in high-redshift galaxies. The presence of a background UV flux strongly influences the formation of molecules by photodissociating them. We explore the evolution after a major merger, leading to the formation of a binary disk. These disks have gas masses of $\rm \sim10^{7}~M_{\odot}$ at a redshift of 5.4. Each disk lies in a separate subhalo as a result of the merger event. The disks are supported by turbulent pressure due to the highly supersonic turbulence present in the halo. For values of $\rm J_{21}=1000$ (internal flux), we find that fragmentation may be enhanced due to thermal instabilities in the hot gas. The presence of dust does not significantly reduce the Lyman alpha emission. The emission of Lyman alpha is extended and originates from the envelope of the halo due to line trapping effects. We also find that dust masses of a few $\rm \times 10^{8}~M_{\odot}$ are required to observe the dust continuum emission from z $\sim 5$ galaxies with ALMA.}
% We noted that Lyman alpha flux of $\sim 10^{-13}~erg/cm^{2}/s$ emerges from $\rm 2 \times 10^{10}~M_{\odot}$ in the presence of dust at z=5.4. Lyman alpha photons scatter with gas clouds due to inhomogeneous distribution of dust which leads to efficient escape of Lyman alpha photons. They are formed despite the strength of the background UV flux in our simulations.We estimate the Lyman alpha emission emanating from the halo. In the future, the dust abundance and evolution of high-redshift galaxies may be further constrained with ALMA.
% context   study the influence of dust on the thermal evolution of protogalactic halos and the assembly of the first galaxies in the presence of a background UV flux
% early type galaxies the depending upon the dust temperature  It is due to the underlying dark matter dynamics in the halo. We also found that dust masses of few $\rm \times 10^{8}~M_{\odot}$ will be required to observe the dust continuum emission from z $\sim 5$ galaxies with ALMA
% aims

%  results

% conclusion 


\maketitle

\section{Introduction}
The observational frontier of high-redshift galaxies is shifting rapidly \citep{2010ApJ...709L.133B,2010MNRAS.409..855B,2011Natur.469..504B,2011ApJ...730L..35V}. High-redshift galaxies have been detected using the spectral imprint of neutral hydrogen. These first galaxies produce copious amounts of Lyman alpha radiation and many have now been detected up to z=7 \citep{2000ApJ...532..170S,2004AJ....128..569M,2006ApJ...648...54S,2008ApJ...675.1076S,2009ApJ...693.1579Y,2009ApJ...696.1164O,2010Natur.467..940L}.

From a theoretical point of view, the first galaxies are formed in atomic cooling halos with $\rm T_{vir} \geq 10^{4}~K$ \citep{2011arXiv1102.4638B,2011arXiv1105.5701J}. Gas in atomic cooling halos may cool through Lyman alpha radiation down to 8000 K. Molecular hydrogen is the only efficient coolant in the absence of metals that can cool the gas down to a few 100 K. Trace amounts of $\rm H_{2}$ can be formed through gas phase reactions \citep{1967Natur.216..976S,1984ApJ...280..465L,1998A&A...335..403G}. HD is another molecule that is a very efficient coolant around 100 K and may cool the gas down close to the CMB temperature. The presence of deuterium creates a small dipole moment, which allows rotational transitions to take place with an appreciable Einstein A coefficient. Despite the fact that $\rm H_{2}$ and HD are the efficient coolants, they are very fragile to Lyman-Werner UV radiation \citep{2008MNRAS.391.1961D,2011A&A...532A..66L,2010MNRAS.402.1249S,2011MNRAS.412.2603W}.

The first galaxies were preceded by the very first stars, forming in minihalos at redshift $\rm \sim 20$ \citep{2002Sci...295...93A,2004ARA&A..42...79B, 2008Sci...321..669Y, 2011Sci...331.1040C, 2011ApJ...727..110C}. These stars emit UV radiation, photoionize gas and photodissociate molecules. Consequently, they influence the subsequent formation of structure in the surrounding intergalactic medium \citep{2007ApJ...665...85J,2011arXiv1105.5701J,2011MNRAS.414.1145M}. It is likely that atomic cooling halos are polluted by trace amounts of metals through Pop III supernova explosions \citep{2010ApJ...716..510G}. The presence of metals or dust significantly influences the thermal evolution of gas and fragmentation becomes inevitable \citep{2001MNRAS.328..969B,2002ApJ...571...30S,2003Natur.422..869S,2003Natur.425..812B,2005ApJ...626..627O,2006ApJ...643...26S,2008MNRAS.385.1443S}. Additional metals may originate from the local presence of stellar winds. The impact of dust and metals in the presence of a background UV flux has been studied by \cite{2008ApJ...686..801O}.

The formation of $\rm H_{2}$ and HD is significantly boosted via grain surface reactions as compared to the gas phase route in the presence of trace amounts of dust (i.e., $\rm > 10^{-5}$ solar) \citep{2004ApJ...611...40C,2009A&A...496..365C}. Dust grains also attenuate UV radiation. The fate of the halo further depends on the content of dust, intensity of the UV radiation field, gas-grain collisional coupling, and the formation of molecules under these conditions \citep{2009ApJ...696.1065J,2009ApJ...694.1161J,2010MNRAS.402..429S}. At high densities, gas-grain collisions have a significant impact on the thermal evolution and the stellar initial mass function \citep{2000ApJ...538..115S,2011ApJ...729L...3D}. These are therefore the key parameters to study.

Although the number of observed Lyman alpha emitters has increased manifold during the past years, the inception of Lyman alpha emitters is not yet fully understood. In our earlier studies \citep{2011MNRAS.413L..33L,2011A&A...532A..66L}, we focused on metal free halos. However, some observations suggest the presence of dusty Lyman alpha emitters \citep{2009Natur.459...61T,2010MNRAS.402.1580O}. The presence of dust has important implications for Lyman alpha emission. It may absorb Lyman alpha photons and re-emit them in the continuum at far-infrared frequencies. \cite{1991ApJ...370L..85N} and \cite{1999ApJ...518..138H} studied the Lyman alpha radiative transfer in a dusty medium. They found that Lyman alpha photons face less attenuation than nonresonantly scattered photons if the medium is inhomogeneous.

In this paper, we study the impact of dust on the dynamical evolution of a protogalactic halo, the formation of binary disks, and the implications of dust for the emission of Lyman alpha photons. We intend to asses how the presence of dust influences the formation of $\rm H_{2}$ and HD molecules for different strengths of the impinging UV radiation field. To complete this task, we have performed state-of-the-art high resolution cosmological simulations by extending our previous chemical model \citep{2011A&A...532A..66L} to include dust physics. The main ingredients of our extended chemical model include the formation of $\rm H_{2}$ and HD molecules on dust grains, extinction of UV radiation by dust grains and gas-grain collisional coupling. We study the evolution for different dust-to-gas ratios and various strengths of the UV radiation field.

Our paper is organized in the following way. In the 2nd section, we describe the numerical schemes and simulation setup. We briefly summarize our chemical model in the 3rd section. We present the results obtained in the 4th section. In the 5th section, we discuss our conclusions.

\section {Computational Methods}

Our numerical simulations have been performed using the extended version of the FLASH code \citep{Dubey2009512}. FLASH is an adaptive mesh refinement (AMR), modular, parallel, grid based code. It can run on massively parallel systems and can be used for a wide variety of astrophysical problems. It uses the message passing interface (MPI) to achieve portability and scalability on different systems. It discretizes the computational domain into nested grid cells by making use of the PARAMESH library. It has two exchangeable grids, a uniform grid and an oct-tree based adaptive grid. We exploit the AMR technique to add resolution in the domain of interest. We capitalize the AMR method and add 15 additional levels of refinement. Consequently, we obtain an effective resolution of 5 pc. We resolve the Jeans length by at least 20 cells to make sure that all essential processes are resolved properly \citep{2011ApJ...731...62F}. This also ensures the fulfillment of Truelove criterion \citep{1997ApJ...489L.179T}. We use an unsplit hydro solver with a 3rd order piece-wise parabolic (PPM) method for hydrodynamic calculations. A multigrid Poisson solver is employed for self-gravity computations. The particle mesh method is used for the  dynamical evolution of a dark matter.

We perform 3-dimensional cosmological simulations and our computational box has a comoving size of 5 Mpc in each dimension. Periodic boundary conditions have been used both for the hydrodynamics and gravity. Our simulations start with cosmological initial conditions and make use of the COSMIC package developed by \cite{1995astro.ph..6070B} to produce Gaussian random field initial conditions. We start our simulations at redshift 90 as computed by the COSMIC code, employing $\rm 3.2 \times 10^{6}$ particles for dark matter gravity calculations. The dark matter particles are distributed according to the initial distribution of the baryons (see below). We use constraint realization, available in the grafic code (part of the COSMIC package), to select a massive halo at the center of a box. We start with an effective grid resolution of $\rm 512^{3}$ in the central 1 Mpc region and set the rest of the box to a resolution of $\rm 128^{3}$ grid cells. In this way, the resolution in the central Mpc is increased by a factor of 4. Our simulations are based on a $\Lambda CDM$ cosmology with WMAP 5-years parameters $\rm \Omega_{m} =0.2581$, $\rm H_{0}=72~km~s^{-1}~Mpc^{-1}$, $\rm \Omega_{b}=0.0441$, assuming a scale invariant power spectrum with $\rm \sigma_{8}=0.8$. The size of the dark matter halo at z=5.4 is $\rm \sim$10 kpc and has a mass of $\rm \sim 2 \times 10^{10}~M_{\odot}$. Simulation results are partly analyzed using YT, i.e., a visualization toolkit for astrophysical data \citep{2011ApJS..192....9T}.

\section{Chemical Model}

The gas in the protogalactic halos is heated up to their virial temperatures where cooling due to Lyman alpha radiation becomes important. HI can cool the gas down to 8000 K. In the absence of metals, $\rm H_{2}$ is the only molecule that can cool the gas down to a few $\rm \times 100$ K. Trace amounts of $\rm H_{2}$ can be formed through the residual fraction of free electrons left over by the recombination epoch. $\rm H_{2}$ can also be formed in shock heated regions \citep{1997ApJ...474....1T,1997NewA....2..209A}. The gas phase formation is dominated by
\begin{equation}
\rm H + e^{-} \rightarrow H^{-} + \gamma .
\label{h2}
\end{equation}
This leads to H$_2$ formation via the reaction
\begin{equation}
\rm H + H^{-} \rightarrow  H_{2} + e^-.
\label{h21}
\end{equation}
HD is a molecule whose cooling becomes efficient below 200 K. The first excited state of HD can be excited around 150 K. It can cool the gas down to about $\rm T_{CMB}=2.75(1+z)$. The main route for HD formation is
\begin{equation}
\rm H_{2} + D^{+} \rightarrow  HD + H^{+}.
\label{h22}
\end{equation}
In a previous study \citep{2011A&A...532A..66L}, we devised a chemical model for a gas of primordial composition, consisting of 36 reactions (23 collisional and 13 radiative). For details, see the appendix of \citep{2011A&A...532A..66L}. We solve the rate equations of the following 12 species: $\rm H,~H^{+},~He,~He^{+},~He^{++},~e^{-},~H^{-},~H_{2},~H_{2}^{+},~D,~D^{+},~and~HD$. We include non-equilibrium ionization, photoionization, collisional ionization, radiative recombination and photodissociation of the above mentioned species. We also include a comprehensive model for cooling and heating, i.e., photo-ionization heating, collisional excitation cooling, recombination cooling, collisional ionization cooling, bremsstrahlung cooling and photo-dissociation heating. Our model also comprises cooling due to $\rm H_{2}$ and HD molecules as well as a multi-level model for atomic hydrogen line cooling (see appendix: table 2 of \cite{2011A&A...532A..66L}). We also include the self-shielding of $\rm H_{2}$ and HD. 
% Further details of these processes can be found in \cite{2011A&A...532A..66L}.

We have extended our previous chemical model to include dust physics. For the description of the dust abundance, we adopt the dust-to-gas ratio measured in units of the solar value, as suggested by \cite{2011MNRAS.tmp.1743S}. Dust models show that supernovae can efficiently produce dust \citep{2001MNRAS.325..726T,2003ApJ...598..785N,2007MNRAS.378..973B}. Indeed, signatures of supernova dust have been observed in the spectra of a high-redshift quasar \citep{2004Natur.431..533M,2010A&A...523A..85G}. 

The presence of dust efficiently boosts the formation of $\rm H_{2}$ and HD molecules \citep{2004ApJ...611...40C,2009A&A...496..365C} for metallicities as low as $\rm Z/Z_{\odot} =10^{-5}$. In order to include a recipe for the formation of $\rm H_{2}$ and HD molecules on dust surfaces, we adopted the $\rm H_{2}$ and HD formation rates on dust grains from \cite{2009A&A...496..365C}, see equations 12 and 13. The formation efficiency of $\rm H_{2}$ and HD on dust grain surfaces depends on the gas and dust temperatures as well as on the grain size distribution and composition.
% Grains are mostly composed of silicates, amorphous carbon(AC), magnetite and corundum \citep{2001MNRAS.325..726T,2007MNRAS.378..973B}. The nucleation processes combined with accretion result in lognormal grain size distributions \citep{2001MNRAS.325..726T,2003ApJ...598..785N}. The presence of reverse shocks in supernovae ejecta destroys a big fraction of dust grains which results in a constant grain size distribution by creating small size dust grains \citep{2007MNRAS.378..973B}. Grain sizes for AC are around 300 $\rm \AA{}$ while for others sizes are around 10-20 $\rm \AA{}$ \citep{2001MNRAS.325..726T}. The grain size distribution in the high-redshift universe differs from the one in the Milky Way.
% Our aim here is to model the formation of $\rm H_{2}$ and HD on dust grains in the high-redshift universe.
We use the grain size distribution of \cite{2001ApJ...548..296W}. In this study, we further focus on carbon (PAHs, AC) and silicate dust grains. We have grain sizes of 3.5-300 $\rm \AA{}$. We assume that dust abundance scales linearly with overall metallicity and that the total dust grain cross section is equal to the the Milky Way for solar metallicity \citep{2009A&A...496..365C}. In this way, we take into account the contribution of smaller grain sizes which may further enhance the $\rm H_{2}$ and HD formation rates due to an increase in surface area. The dust temperature in the presence of a background UV flux  is computed as \citep{1991ApJ...377..192H,2005A&A...436..397M}
\begin{eqnarray}
\rm T_{d}=\{8.9\times10^{-11} \nu_{0}G_{0}10^{-1.8 A_{\nu}} + T_{CMB}^{5} +3.4 \times 10^{-2}\\ \nonumber
\rm [0.42-log(3.5\times10^{-2}\tau_{100}T_{0})]\times \tau_{100}T_{0}^{-0.2} \}^{0.2}
\label{dust1}
\end{eqnarray} 
where $\rm \nu_{0}=2.65\times10^{15}~s^{-1}$, $\rm G_{0}$ is the far-ultraviolet (FUV) flux in units of the equivalent Habing flux \citep{1968BAN....19..421H}, $\rm T_{0}=12.2G_{0}^{0.2}~K$ is the equilibrium dust temperature and $\rm \tau_{100}=2.7\times 10^{2}/T_{0}^{5}G_{0} Z/Z_{\odot}$ is the optical depth at 100 $\rm \mu m$. The dust temperature is computed assuming that each dust grain is illuminated by attenuated FUV flux, cosmic microwave background radiation (CMB), and the infrared radiation field from warm dust emission \citep{1991ApJ...377..192H}. If gas and dust grains have different temperatures, they can exchange heat through collisions. We compute the gas-grain collisional heating and cooling rates as given in the equation below \citep{2005A&A...436..397M}

\begin{eqnarray}
\rm \Gamma_{coll} =1.2\times10^{-31}n^{2} (Z/Z_{\odot}) \left({T_{k} \over 1000}\right)^{1/2} \left({100\AA{} \over a_{min}} \right)^{1/2} \\ \nonumber
\rm \times[1.0-0.8exp(-75/T_{k})](T_{d}-T_{k})
\label{dust33e}
\end{eqnarray}
$\rm T_{d}$ and $\rm T_{k}$ are dust and gas temperatures, respectively, n is the gas number density, and $\rm a_{min}$ is the minimum dust grain size. We set the minimum grain size to 3.5 $\rm \AA{}$ \citep{2005A&A...436..397M}. Grain surface reactions are often more effective than their gas-phase counterparts. We have included $\rm H_{2}$, HD, H, D and He formation on the surfaces of dust grains in our chemical model. The rates for these reactions are taken from \cite{2007ApJ...666....1G} and \cite{2009A&A...496..365C}, and are listed in table \ref{Table1}. We have also included the $\rm H_{2}$ collisional dissociation cooling, photoelectric heating by UV irradiated dust grains, gas-phase $\rm H_{2}$ formation heating, and heating by $\rm H_{2}$ formation on dust grains. These are listed in table 7 of \cite{2007ApJ...666....1G}.

\begin{table*}
 \centering
\caption{~~~~~~~~~~~~~~~~~~~~~~~~~~~~~ \textbf{Collisional and radiative rates}}
\begin{tabular}{llll}

\hline\hline
No. & Reaction & Rate & Reference \\ \hline
1 & $\rm H_{2}$ formation rate on dust grains  &  \textbf{see reference} &     CS09  \\ 
2 & HD formation rate on dust grains  &  \textbf{see reference}  &     CS09  \\ 
3 &  \HIId ~+ \ed $\rightarrow$   \HId  & \textbf{see reference} &     GJ07  \\
4 &  \DIId ~+ \ed $\rightarrow$ \DId  & \textbf{see reference} &     GJ07  \\
5 &  \HeIId ~+ \ed $\rightarrow$ \HeId & \textbf{see reference} &     GJ07  \\
\hline 
\end{tabular}
\label{Table1}

\flushleft 
Glover and Jappsen (2007);CS09 :Cazaux and Spaans 2009.
\end{table*}

During the epoch of reionization, a background UV flux is produced by stellar populations. The background UV flux can photoheat and photoionize the gas, and can photodissociate molecules. We have modeled the background UV flux in the following way \citep{2008ApJ...686..801O}
\begin{equation}
\rm J_{\nu}^{ex} = J_{21}10^{-21}[B_{\nu}(T_{*})/B_{\nu H}] f(\nu)~ erg~cm^{-2}sr^{-1}s^{-1}Hz^{-1},
\label{jrad}
\end{equation}
where $\rm J_{21}$ is the background UV radiation field intensity below the Lyman limit and $\rm T_{*}$ is the color temperature of a star. We assume a black body radiation spectrum with $\rm T_{*}=10^{4}~K$. $\rm f(\nu)$ is one for non-ionizing radiation and is equal to $\rm f_{esc}$ for ionizing radiation. $\rm B_{\nu H}$ is the black body spectrum at the Lyman limit ($\rm J_{21}=10^{-21}~erg/cm^{2}/s/Hz/sr$). $\rm B_{\nu }$ is the stellar radiation spectrum (i.e., black body spectrum) as given by
\begin{equation}
\rm B_{\nu}(T) = {2h \nu ^{3} \over c^{2}} {1 \over (exp(h\nu/kT) -1) }.
\label{bnew}
\end{equation}
Here h is Planck's constant, $\nu$ is the frequency of radiation, c is the speed of light, k is Boltzmann's constant and T is the radiation temperature. $\rm J_{IH}=J_{21}\times f_{esc}$ is the ionizing flux, where $\rm f_{esc}$ is again the escape fraction of ionizing radiation. These ionizing photons are attenuated while propagating through intergalactic medium. We have taken this effect into account  in an approximate manner by computing the optical depth for each frequency bin in the following manner
\begin{equation}
\rm \tau_{i}(\nu _{j}) = \Sigma_{i} \sigma_{i}(\nu_{j}) \times n_{i} \times \lambda_{J} ,
\label{taaau}
\end{equation}
where i stands for the specie and j denotes the frequency bin, $\rm \sigma_{i}$ is the photoionization cross section of the specie, n is the gas number density and $\rm \lambda_{J}$ is the local Jeans length. The radiation flux also becomes attenuated by dust grains. These grains both absorb and scatter the radiation. The dust opacity strongly depends on the frequency of the UV flux. We use a fit to the extinction curve by \cite{1991ApJS...77..287R} to compute the dust opacity as given by
\begin{equation}
\rm \tau_{dust}(\nu_{i}) =n_{H}(i)\sigma_{d}(i)A_{\nu}(i) \lambda_{J}\times Z/Z_{\odot}
\label{dustff}
\end{equation}
here $\rm n_{H}$ is the number density of hydrogen, $\rm \sigma_{d}$ is the effective dust cross section, and $\rm A_{\nu}$ is the dust extinction coefficient. Total attenuation of the radiation field is given by
\begin{equation}
\rm J^{att} (\nu) = J^{ext}(\nu) \times exp(-\tau(\nu_{i})) \times exp(-\tau_{dust}(\nu_{i})) .
\label{bnew2}
\end{equation}

The stellar sources formed inside the halos will produce a local UV flux. We have also modeled this flux and studied its influence on the collapse and chemistry of the gas physics. As these sources are formed in the high density regions, we enhance the background UV flux by factors of 100 and 1000 above densities of 1 $\rm cm^{-3}$.
%In rest of this article, we use term gas-to-dust ratio ($\rm D/D_{\odot}$) which is $\rm D/D_{\odot}=\rm Z/Z_{\odot} \times Z_{\odot}\times 0.5$. The radiation flux gets attenuated by the dust grains. These grains also scatter the radiation. Dust opacity strongly depends on the frequency of the UV flux. We use a fit to the extinction curve by \cite{1991ApJS...77..287R} to compute the dust opacity as given by
% 
% \begin{equation}
% \rm \tau_{dust} =n_{H}\sigma_{d}A_{\nu}\lambda_{J}\times Z/Z_{\odot}
% \label{dust2}
% \end{equation}
% here $\rm n_{H}$ is the number density of hydrogen, $\rm \sigma_{d}$ is the effective dust cross section and $\rm A_{\nu}$ is dust extinction coefficient.
\begin{figure*}[htb!]
\centering
\begin{tabular}{c c c}
\begin{minipage}{8cm}
 \hspace{0.4cm}
\includegraphics[scale=0.21]{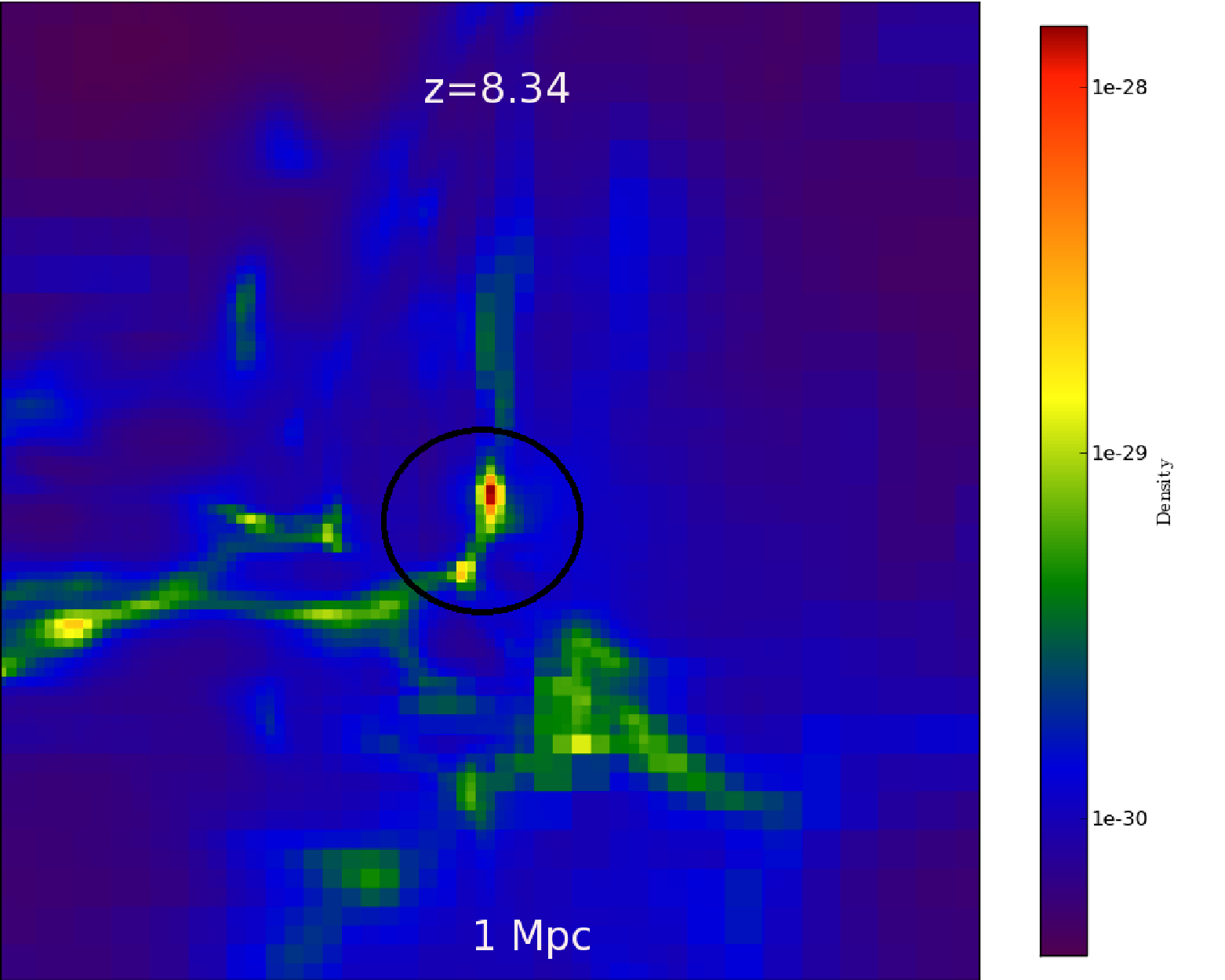}
\end{minipage} &
 \hspace{-5cm}
\begin{minipage}{8cm}
\includegraphics[scale=0.21]{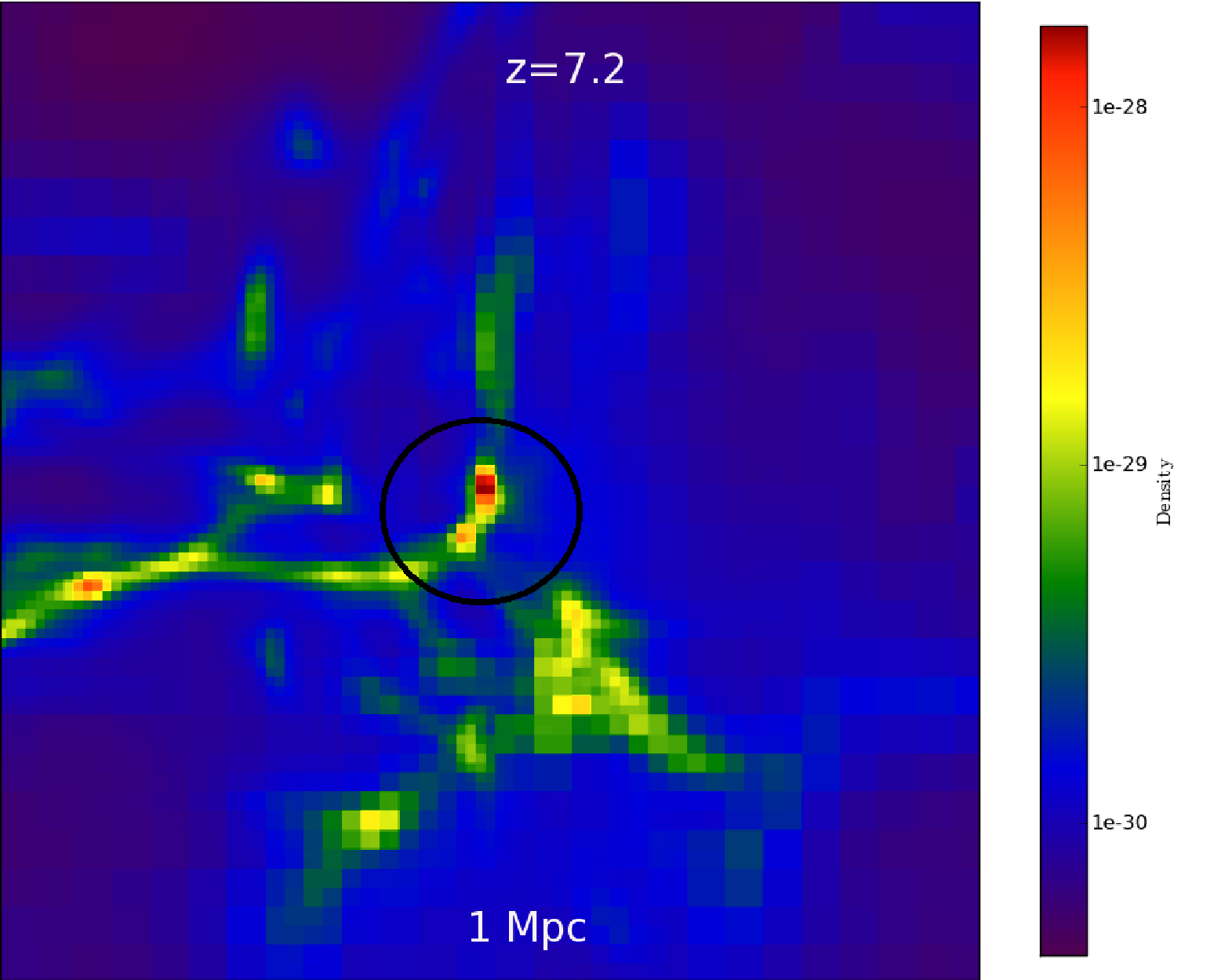}
\end{minipage} &
\begin{minipage}{8cm}
\hspace{-5.4cm}
\includegraphics[scale=0.21]{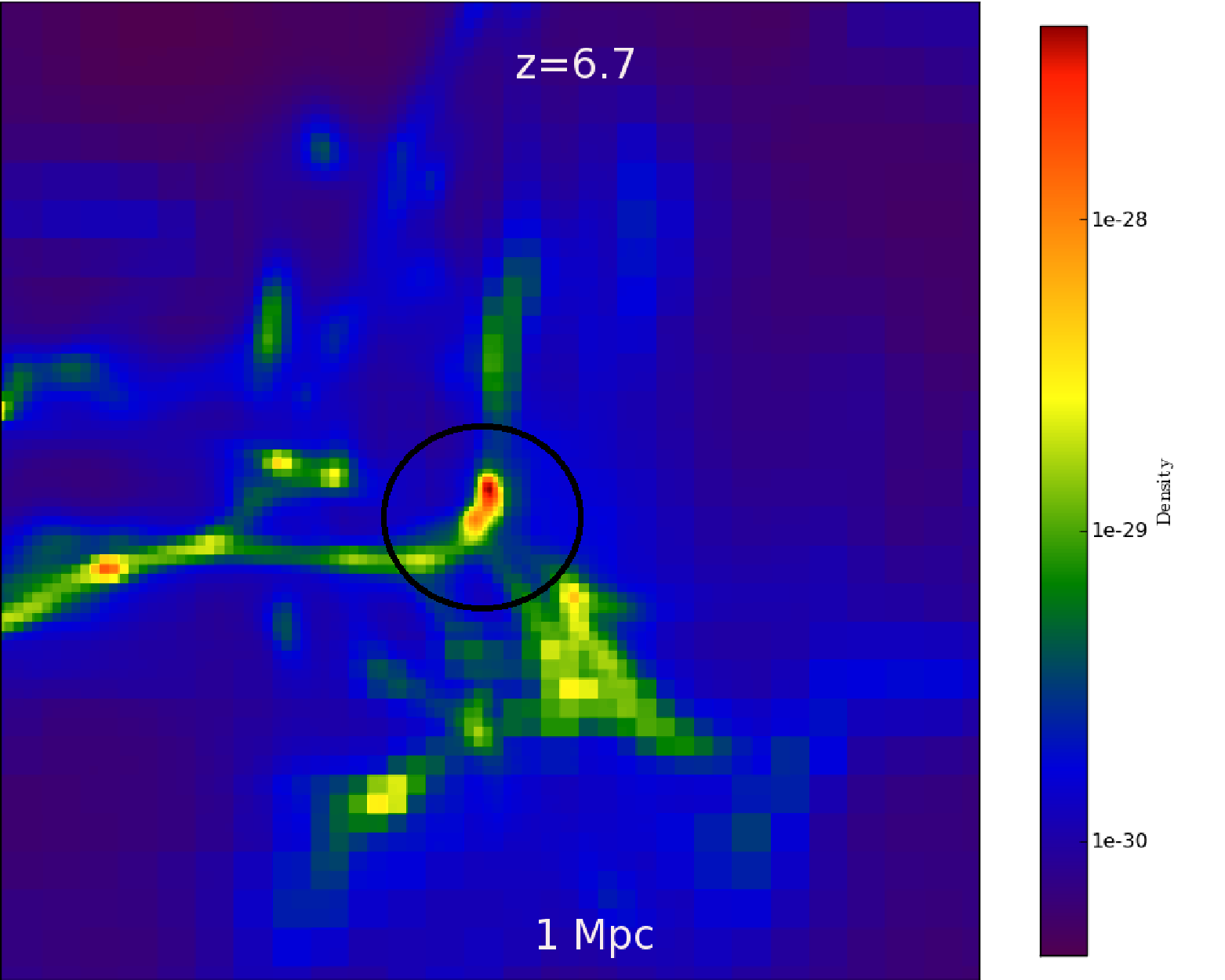}
\end{minipage} \\ \\
\begin{minipage}{8cm}
\hspace{0.4cm}
\includegraphics[scale=0.21]{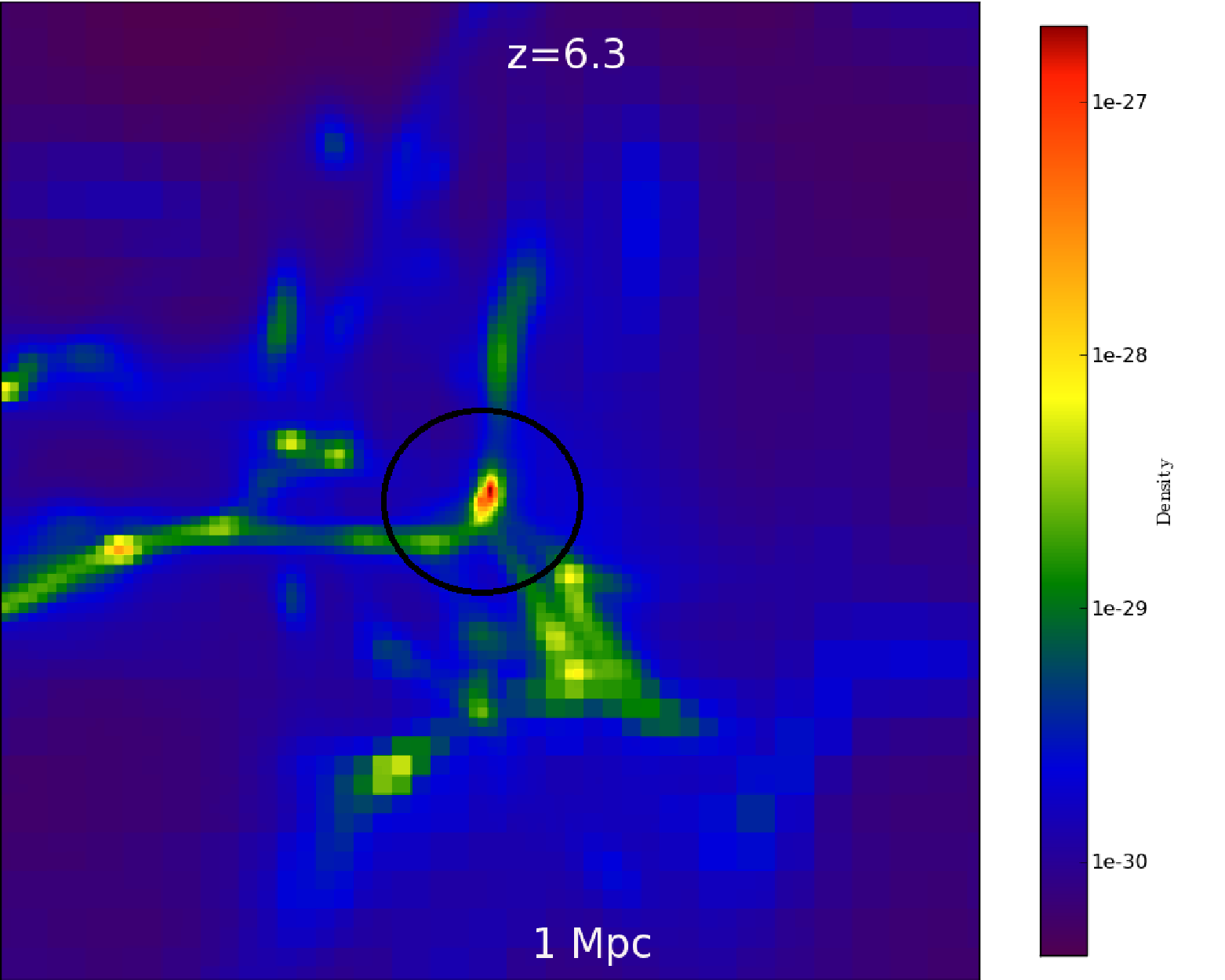}
\end{minipage} &
\begin{minipage}{8cm}
\hspace{-2.5cm} 
\includegraphics[scale=0.21]{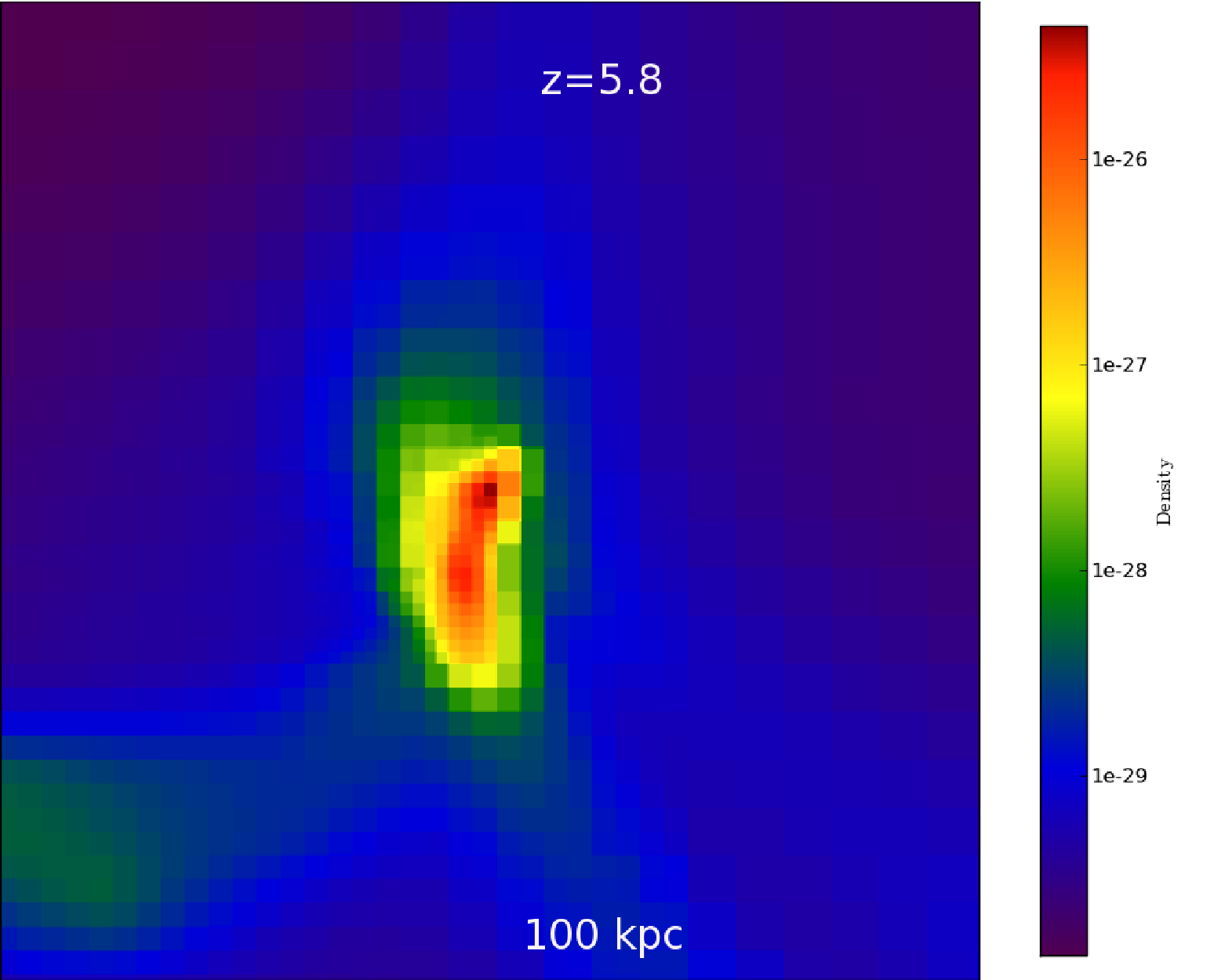}
\end{minipage} &
\begin{minipage}{8cm} 
\hspace{-5.5cm}
\includegraphics[scale=0.21]{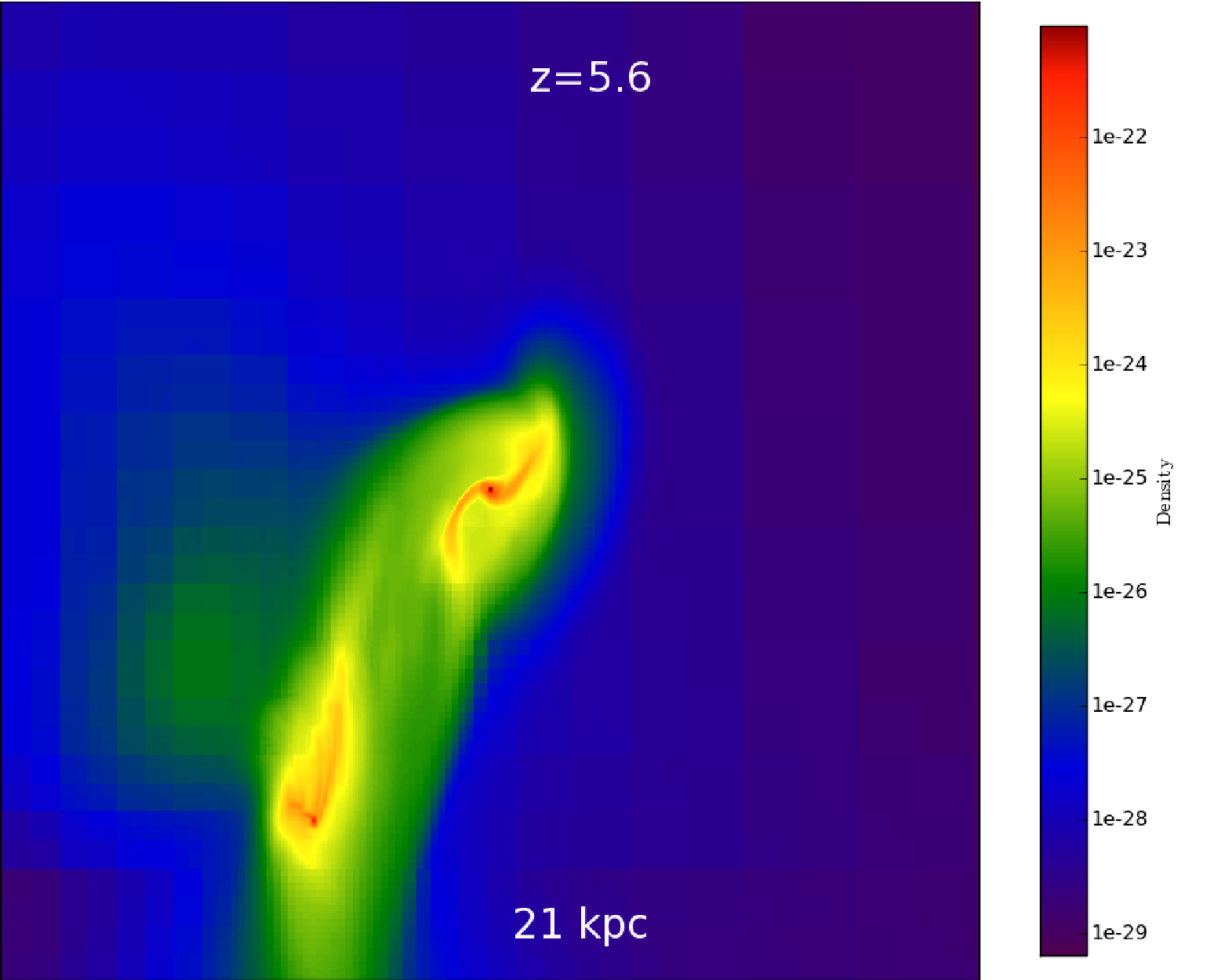}
\end{minipage}
\end{tabular}
\caption{Time sequence of density snapshots at different redshifts. Redshift and distance scales are mentioned in each panel, and shown for model A. The values of density shown in the colorbar are in comoving units [$\rm g/cm^{3}$]. The merging of clumps is shown inside the black circle.}
\label{figured1}
\end{figure*}

\begin{figure*}[htb!]
\centering
\begin{tabular}{c c}
\begin{minipage}{8cm}
% \includegraphics[scale=0.48]{DENS-new1-dust-J21-radmulti.ps}
% \end{minipage} &
% \begin{minipage}{8cm}
% \includegraphics[scale=0.48]{DENS1-new1-dust-J21-radmulti.ps}
\includegraphics[scale=0.28]{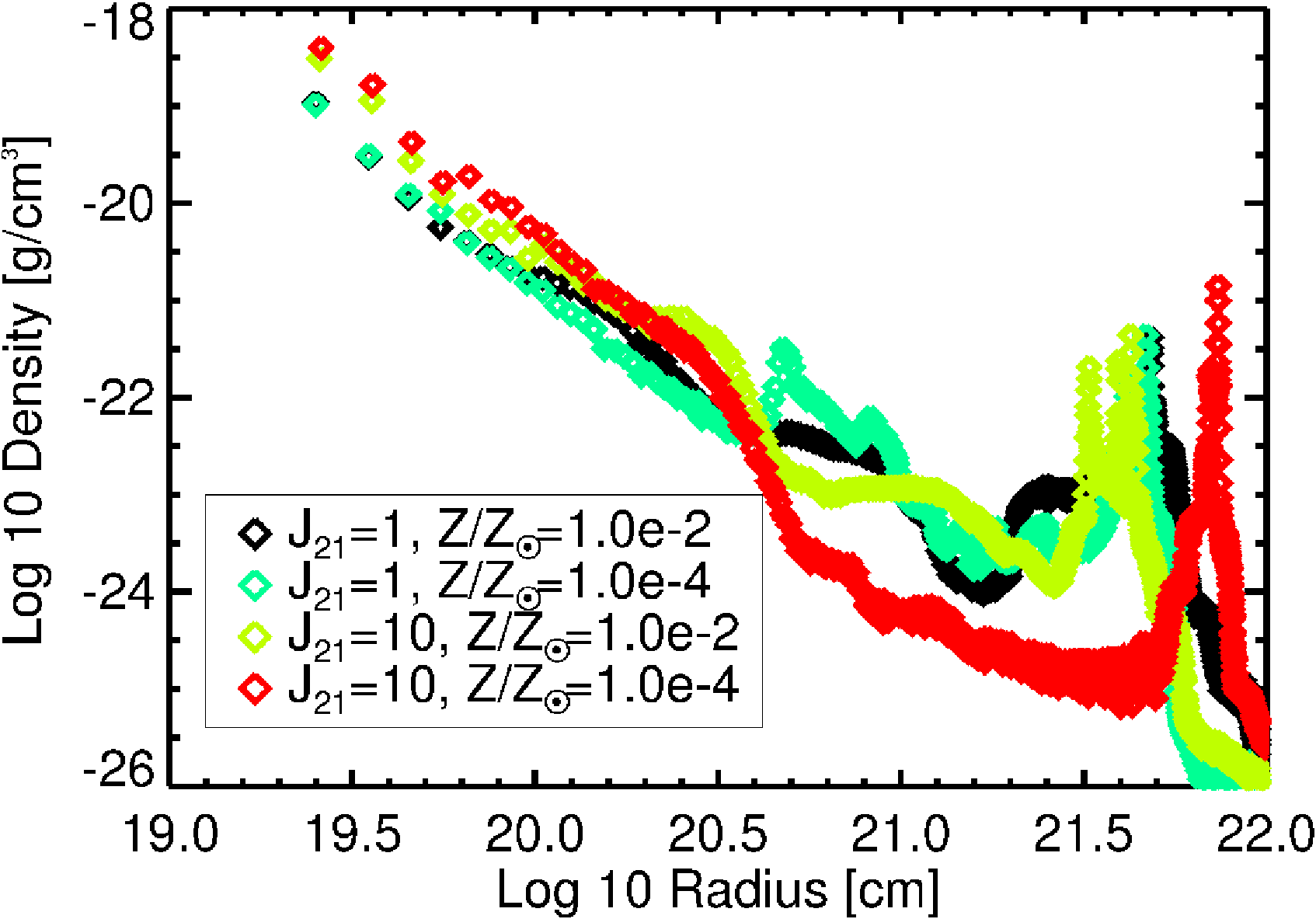}
\end{minipage} &
\begin{minipage}{8cm}
\includegraphics[scale=0.28]{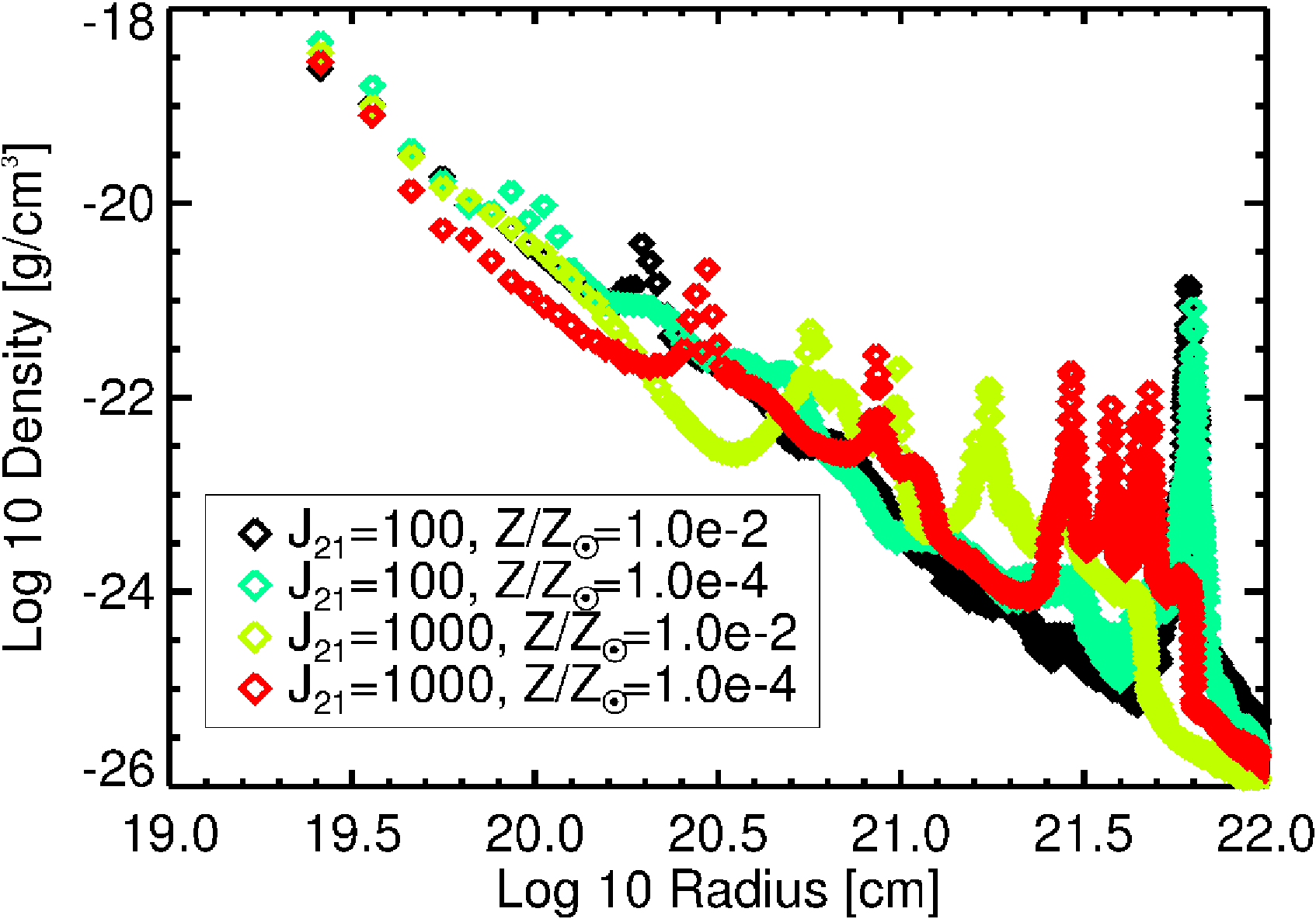}
\end{minipage} 
\end{tabular}
\caption{Average density radial profiles for different strengths of the background UV field and dust-to-gas ratios at z=5.4. The values of the background UV flux and metallicities are shown inside each figure. The right hand side panel is for an internal UV radiation flux (model E, F, G, H) in addition to a background UV flux of $\rm J_{21}=1$. The value of the escape fraction  for ionizing radiation is 10\% for both panels.}
\label{figure1}
\end{figure*}

\begin{figure*}[htb!]
\centering
\begin{tabular}{c c}
\begin{minipage}{8cm}
% \hspace{0.27cm}
% \includegraphics[scale=0.48]{TEMP-new1-dust-J21-radmulti.ps}
% \end{minipage} &
% \begin{minipage}{8cm}
% \includegraphics[scale=0.48]{HII-new1-dust-J21-radmulti.ps}
% \end{minipage} \\  \\
% 
% \begin{minipage}{8cm}
% \includegraphics[scale=0.48]{H2-new1-dust-J21-radmulti.ps}
% \end{minipage} &
% 
% \begin{minipage}{8cm}
% \includegraphics[scale=0.48]{HD-new1-dust-J21-radmulti.ps}
% \end{minipage}
\includegraphics[scale=0.28]{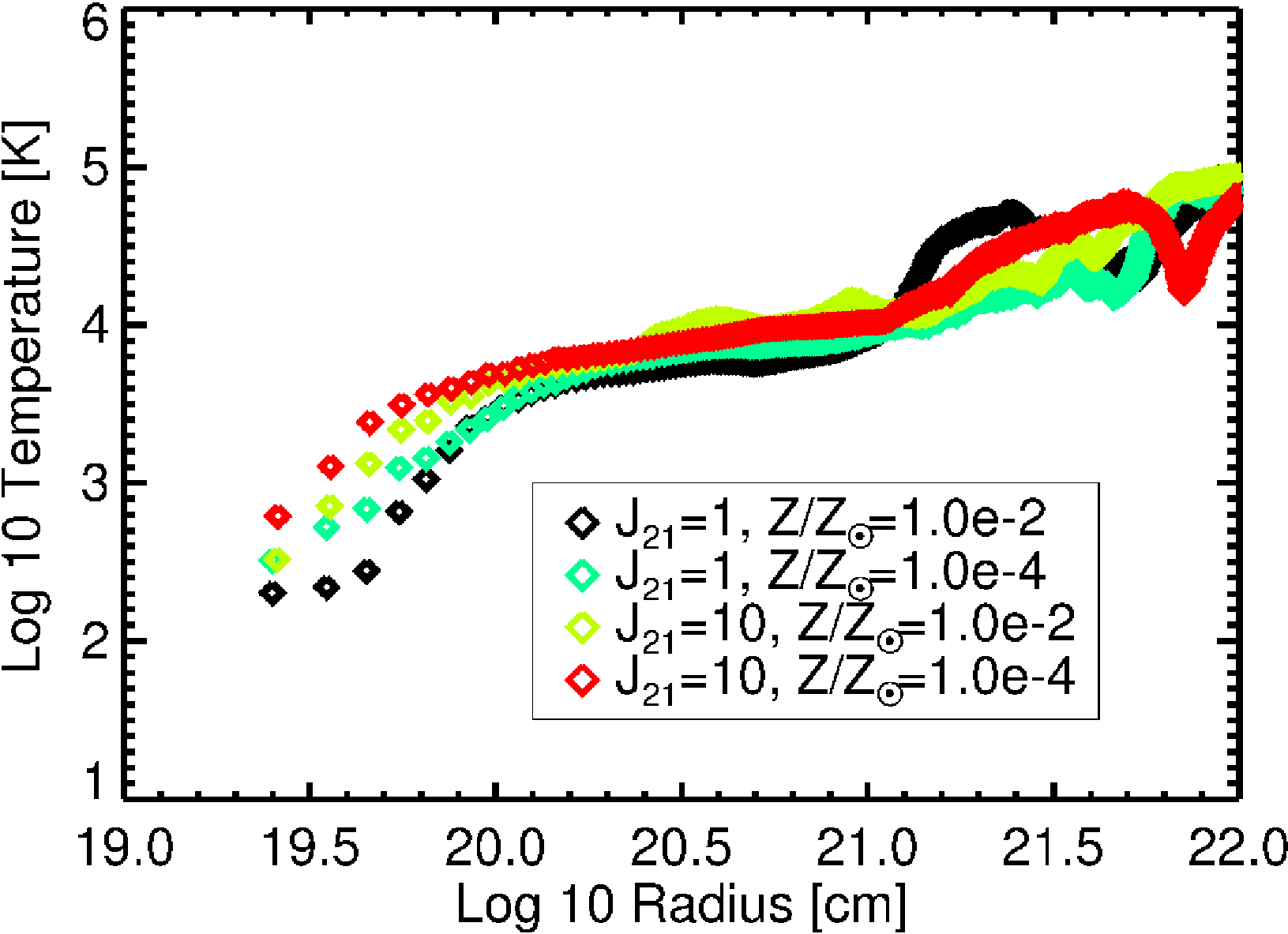}
\end{minipage} &
\begin{minipage}{8cm}
\includegraphics[scale=0.28]{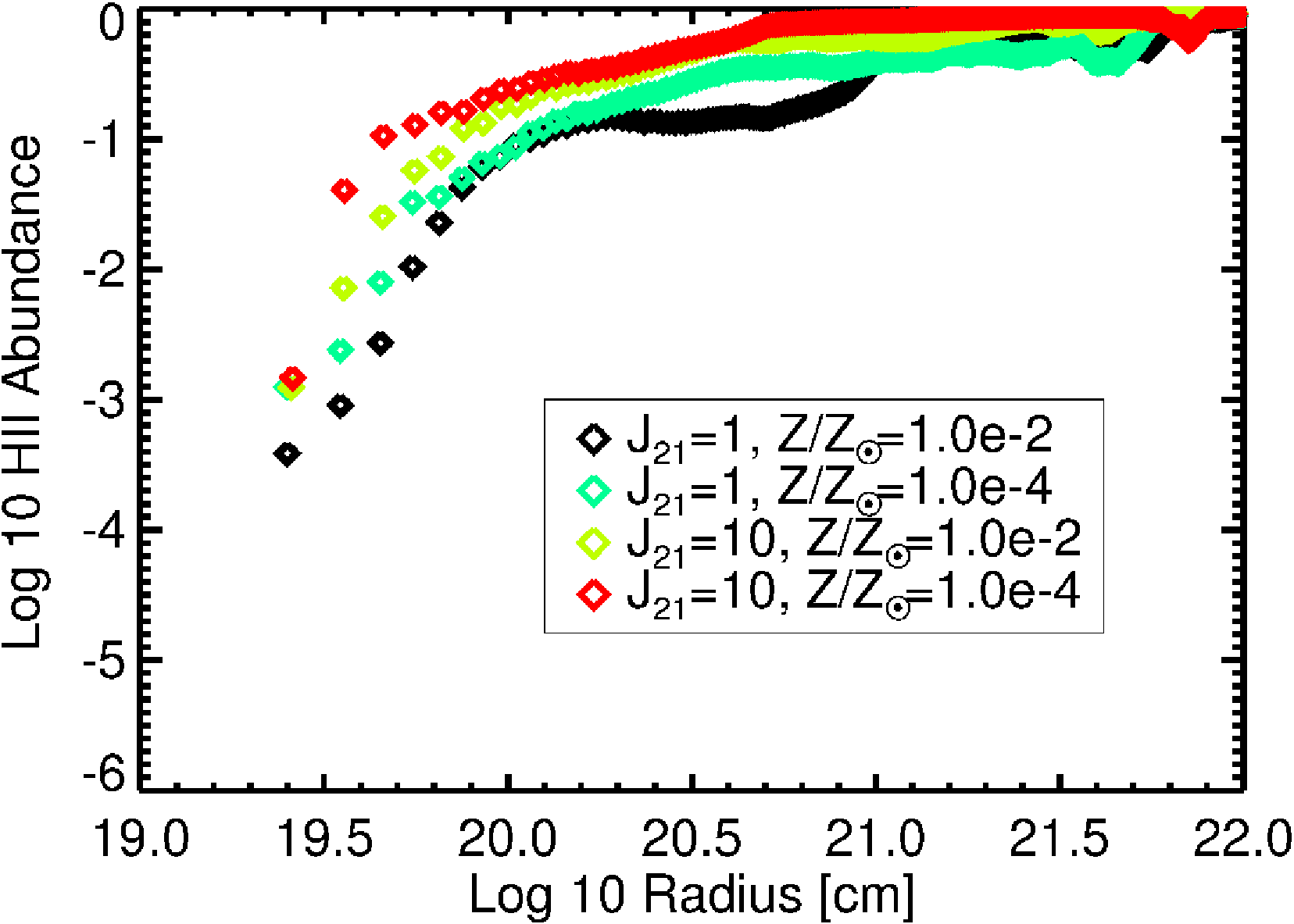}
\end{minipage} \\  \\

\begin{minipage}{8cm}
\includegraphics[scale=0.28]{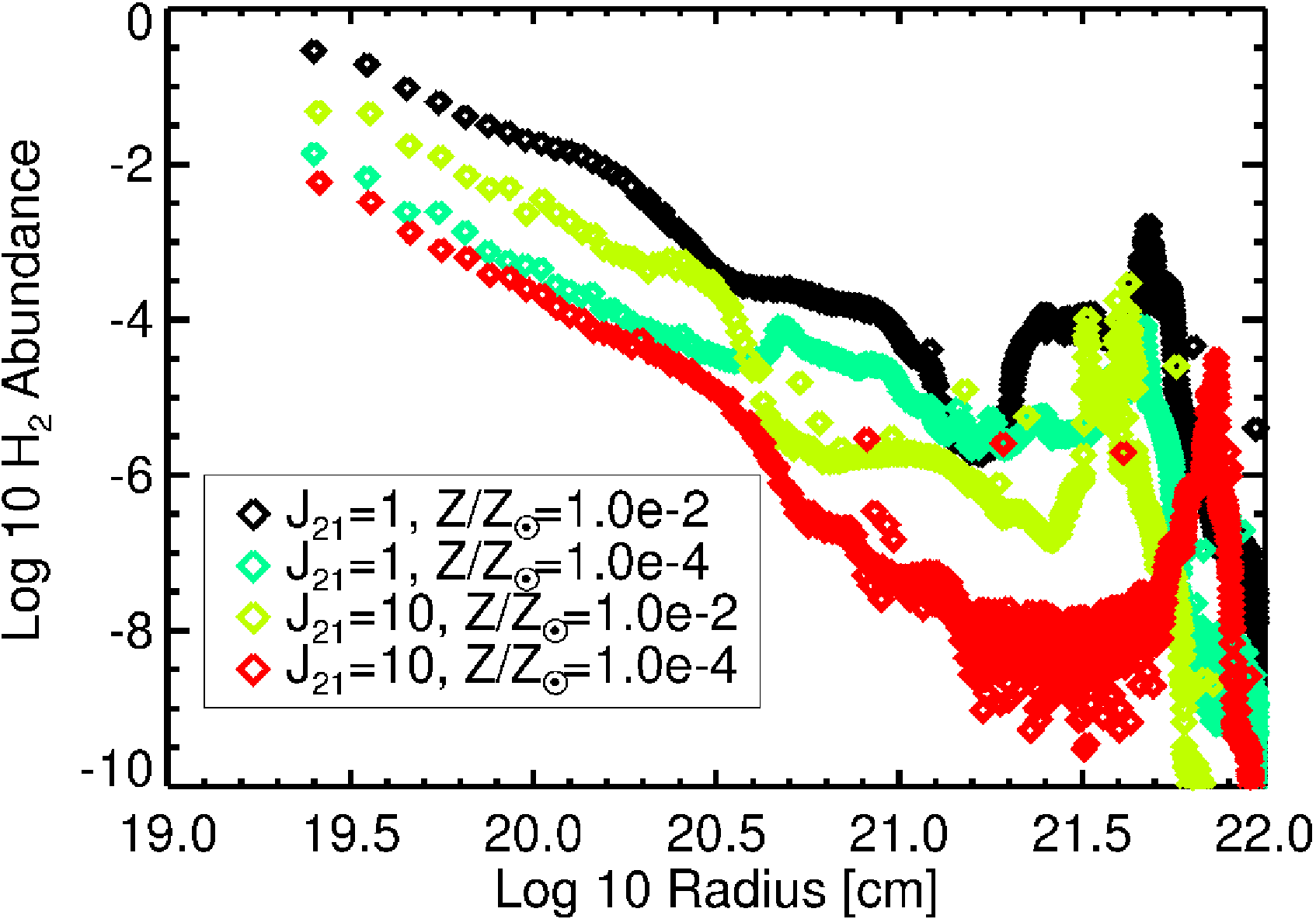}
\end{minipage} &

\begin{minipage}{8cm}
\includegraphics[scale=0.28]{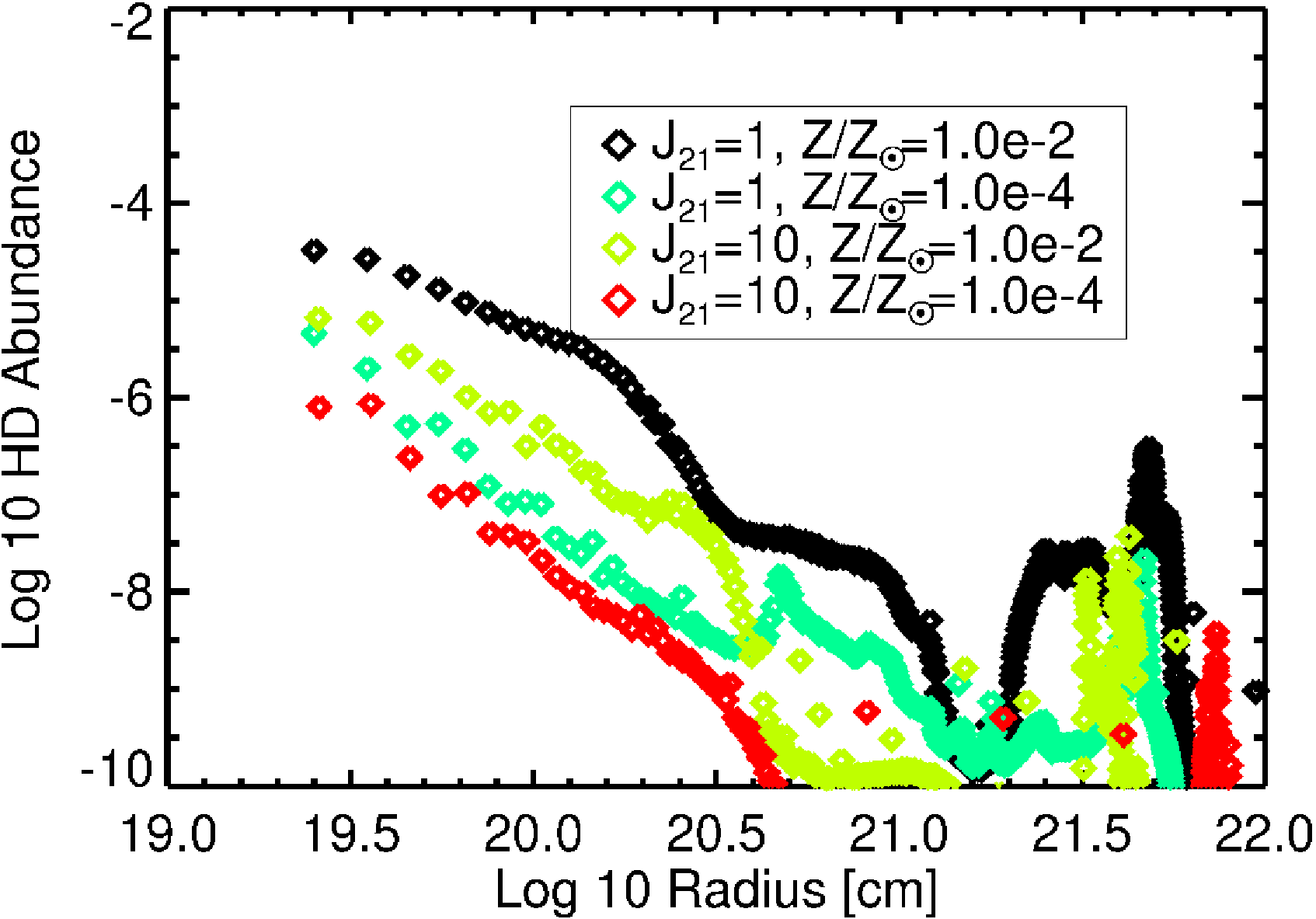}
\end{minipage}
\end{tabular}
\caption{Average radial profiles for different background UV fields and dust-to-gas ratios at z=5.4. The upper left panel of this figure shows the temperature radial profile of the halo. The HII abundance radial profile for the halo is depicted in the upper right panel. $\rm H_{2}$ abundance is shown in the lower left panel. The lower right panel shows the HD radial profile of the halo. The escape fraction of ionizing radiation is 10\% for all the panels shown in the figure.}
\label{figure2}
\end{figure*}

\begin{figure*}[htb!]
\centering
\begin{tabular}{c c}
\begin{minipage}{8cm}
 \hspace{-0.6cm}
\includegraphics[scale=0.2]{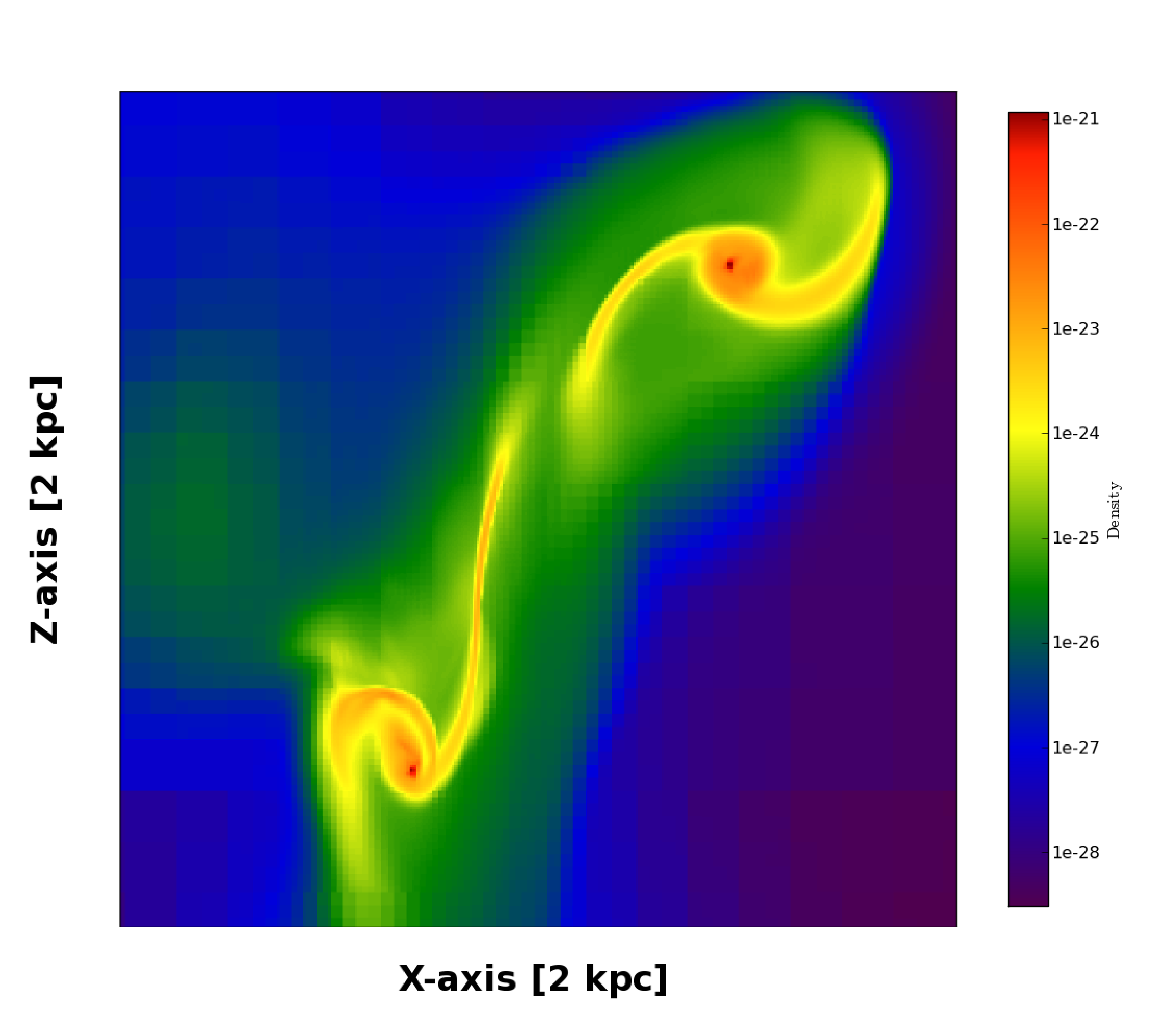}
\end{minipage} &
\begin{minipage}{8cm}
\includegraphics[scale=0.28]{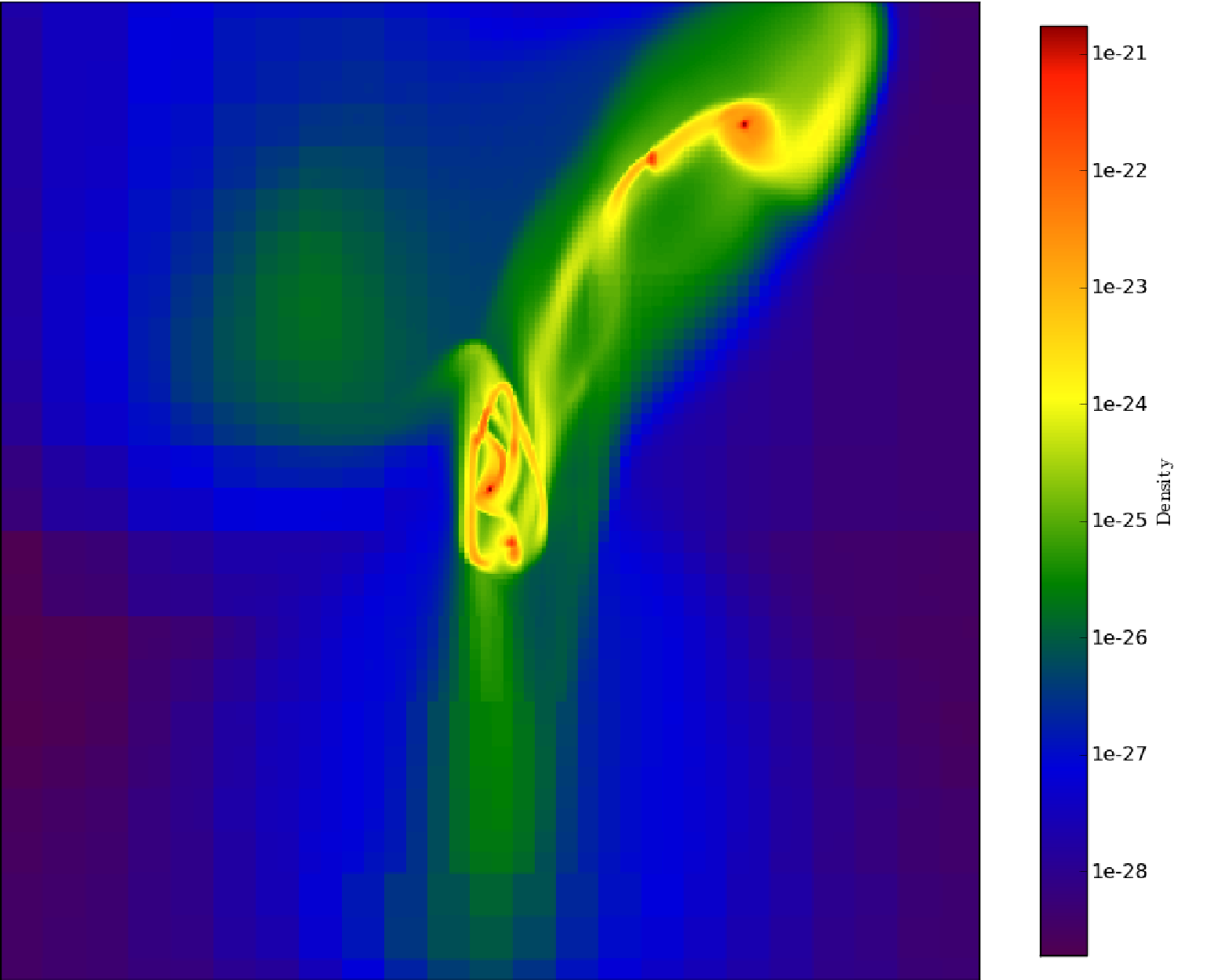}
\end{minipage} \\  \\

\begin{minipage}{8cm}
\includegraphics[scale=0.28]{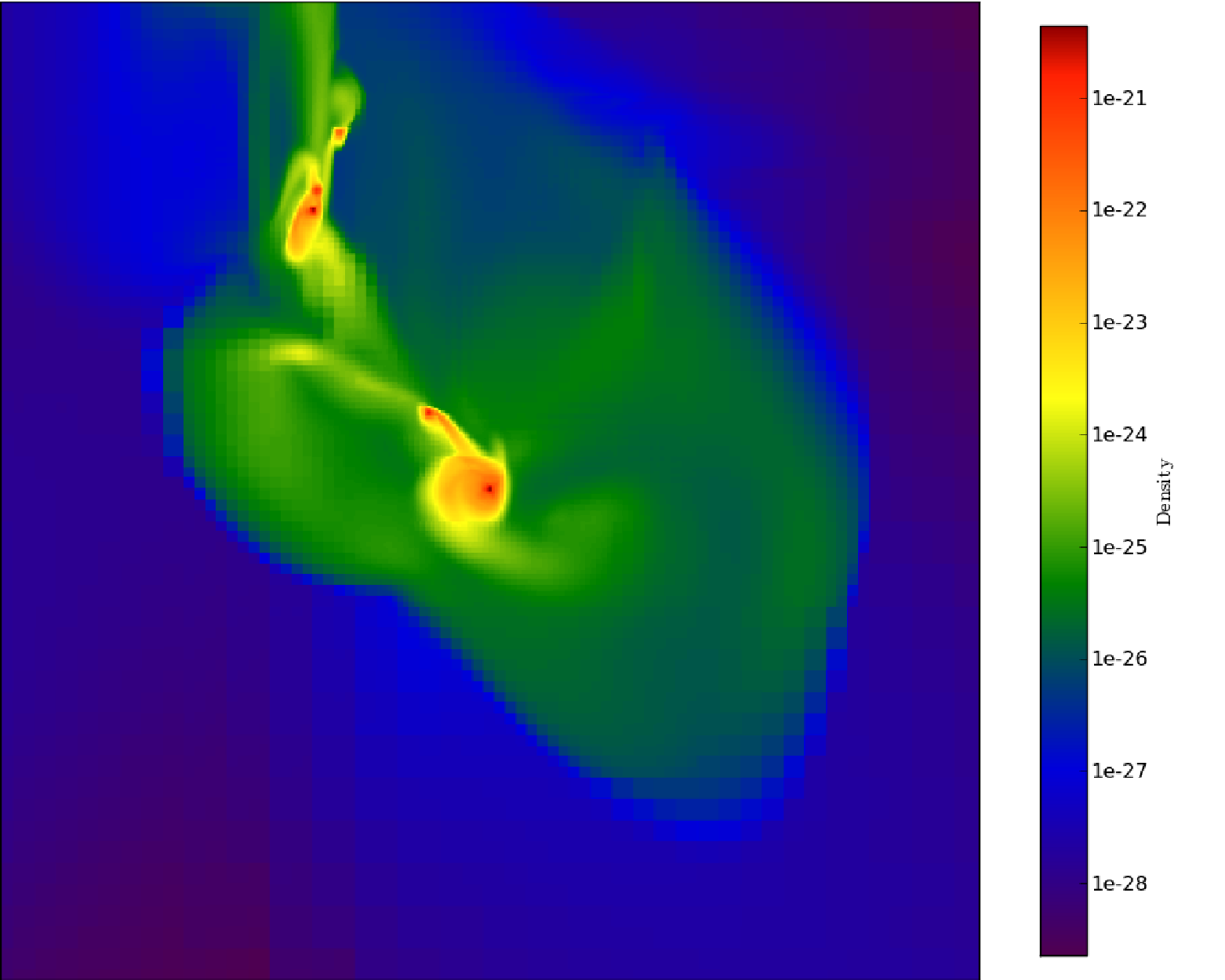}
\end{minipage} &

\begin{minipage}{8cm}
\includegraphics[scale=0.28]{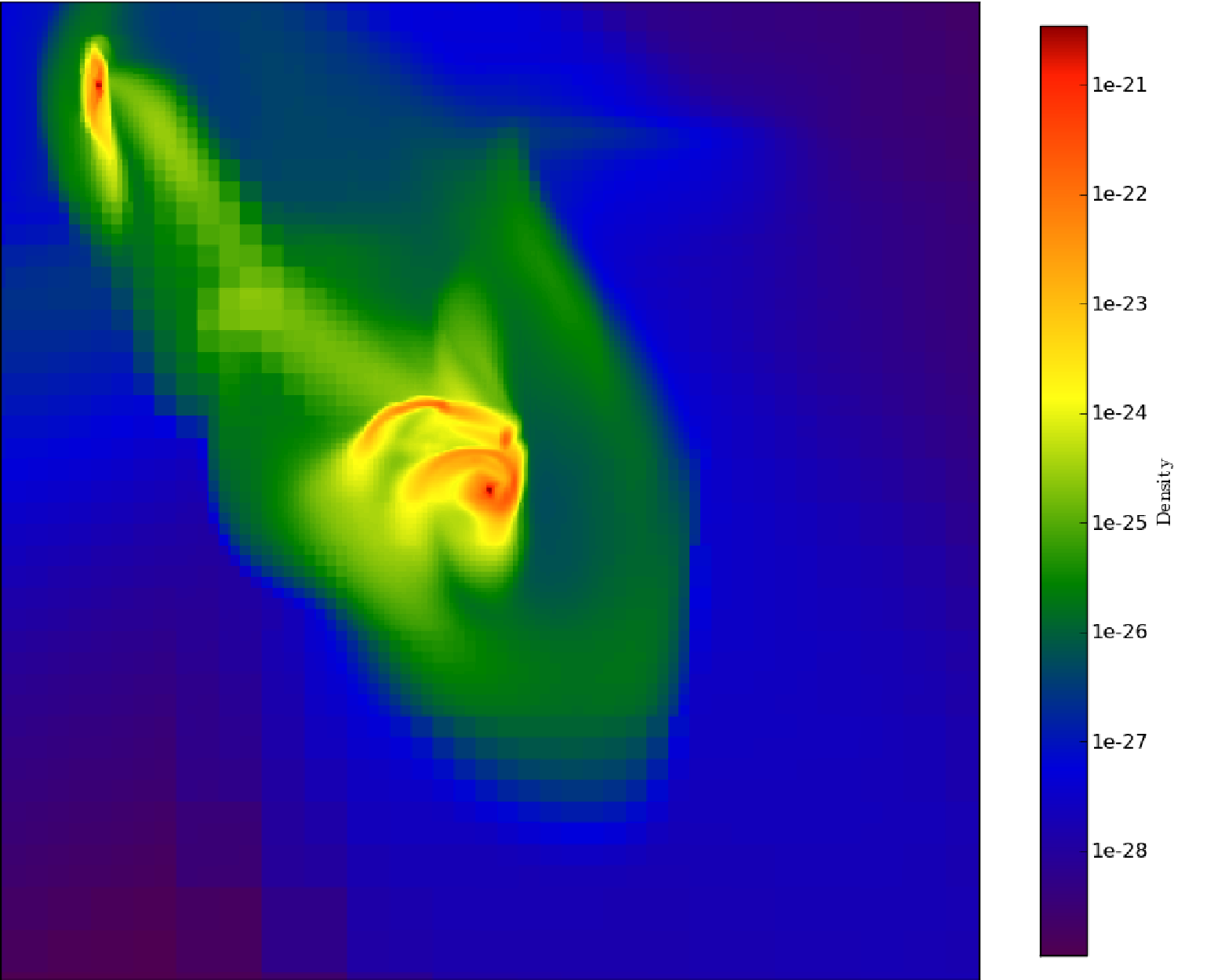}
\end{minipage}

\end{tabular}
\caption{Density weighted projection of gas density for different background UV fields and dust-to-gas ratios at redshift 5.4. The top left panel shows the density for model A. The density structure for model B is shown in the upper right panel. The bottom left panel shows the density structure for model C while the bottom right panel shows the density structure for model D. The escape fraction of ionizing radiation is 10\% for all the panels. The values of density shown in the colorbar are in comoving units [$\rm g/cm^{3}$].}
\label{figured}
\end{figure*}

\begin{figure*}[htb!]
\centering
\begin{tabular}{c c}
\begin{minipage}{8cm}
% \hspace{0.27cm}

 \includegraphics[scale=0.28]{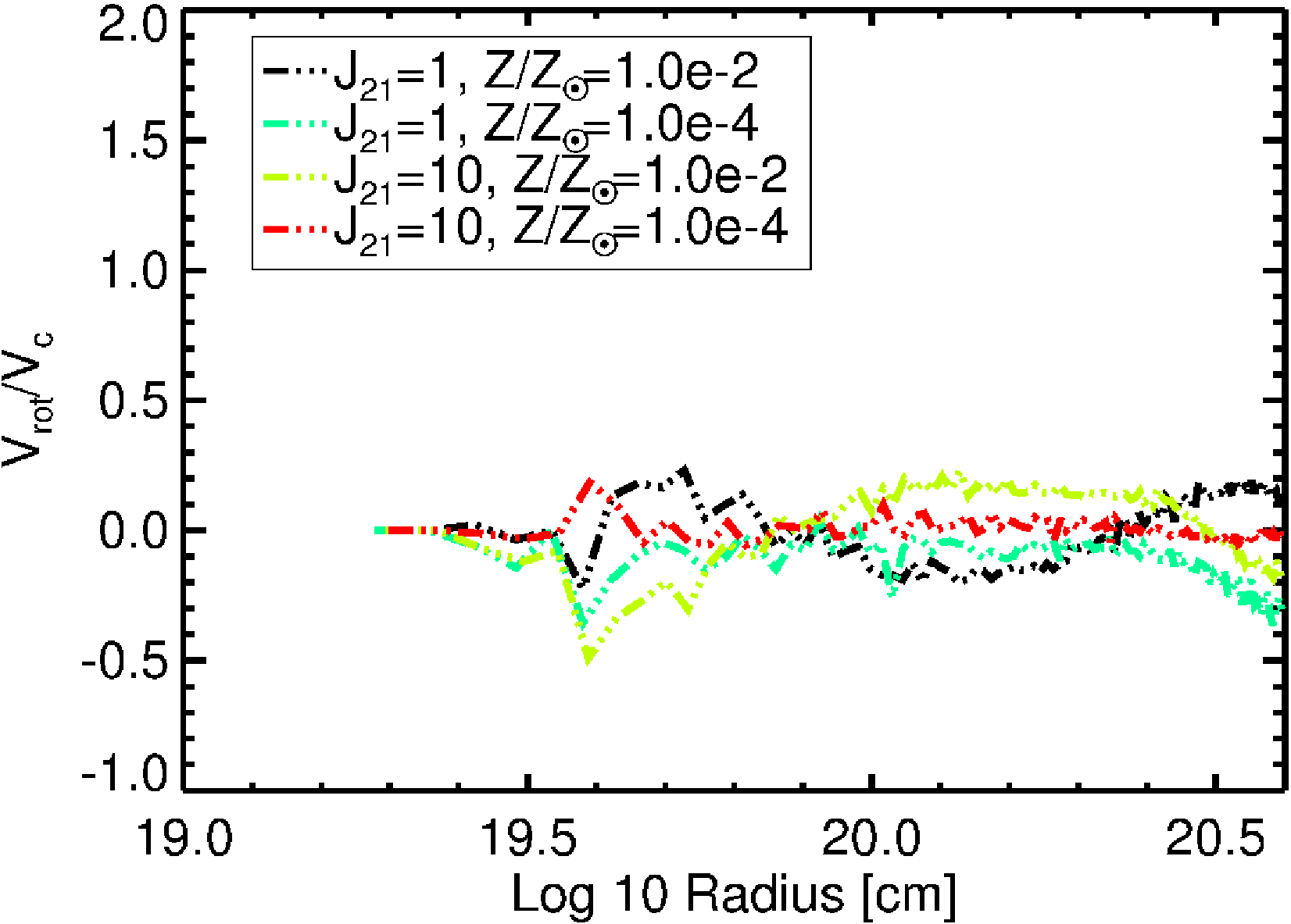}
\end{minipage} &
\begin{minipage}{8cm}
 \includegraphics[scale=0.28]{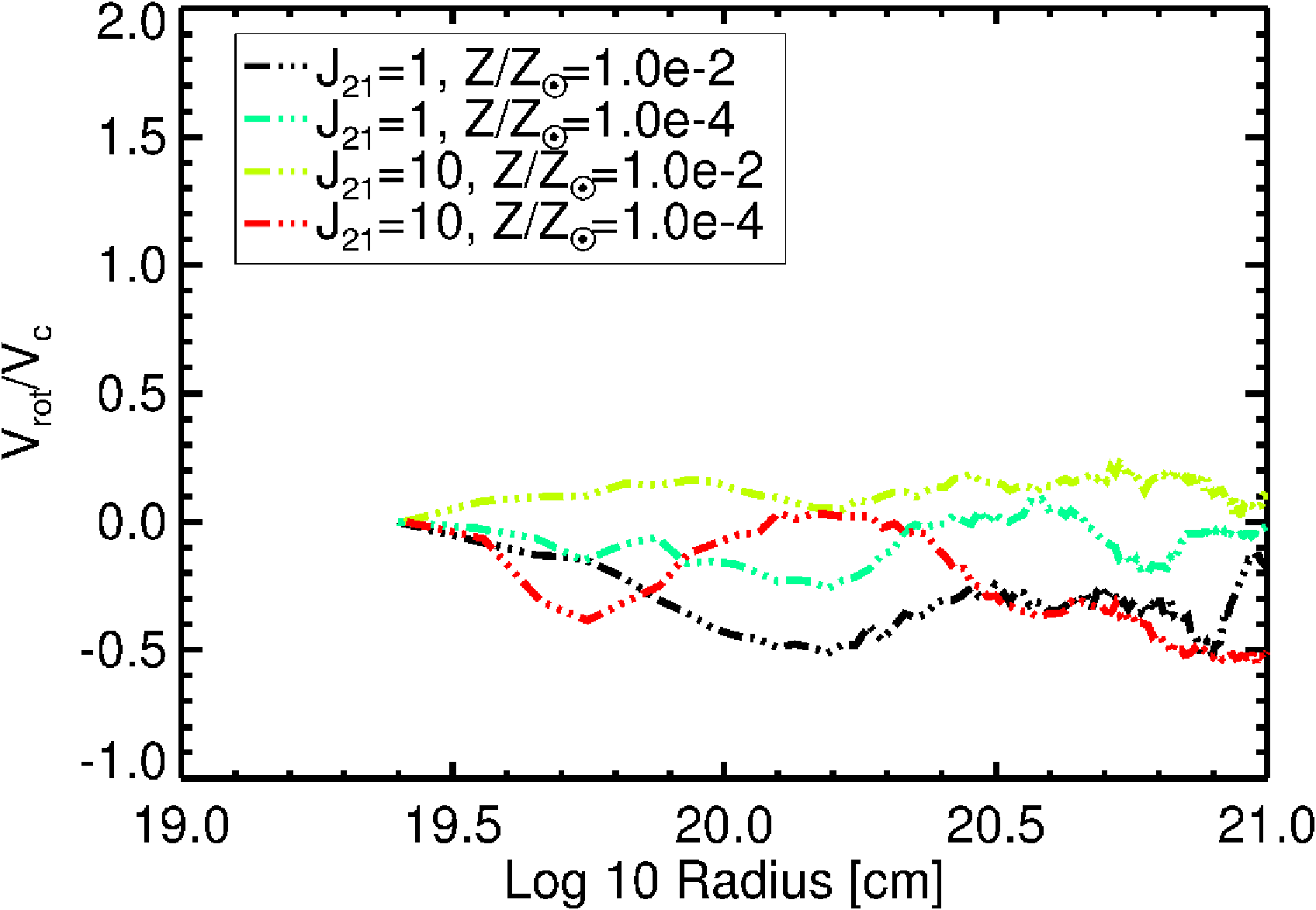}
\end{minipage} \\  \\
 
\begin{minipage}{8cm}
 \includegraphics[scale=0.28]{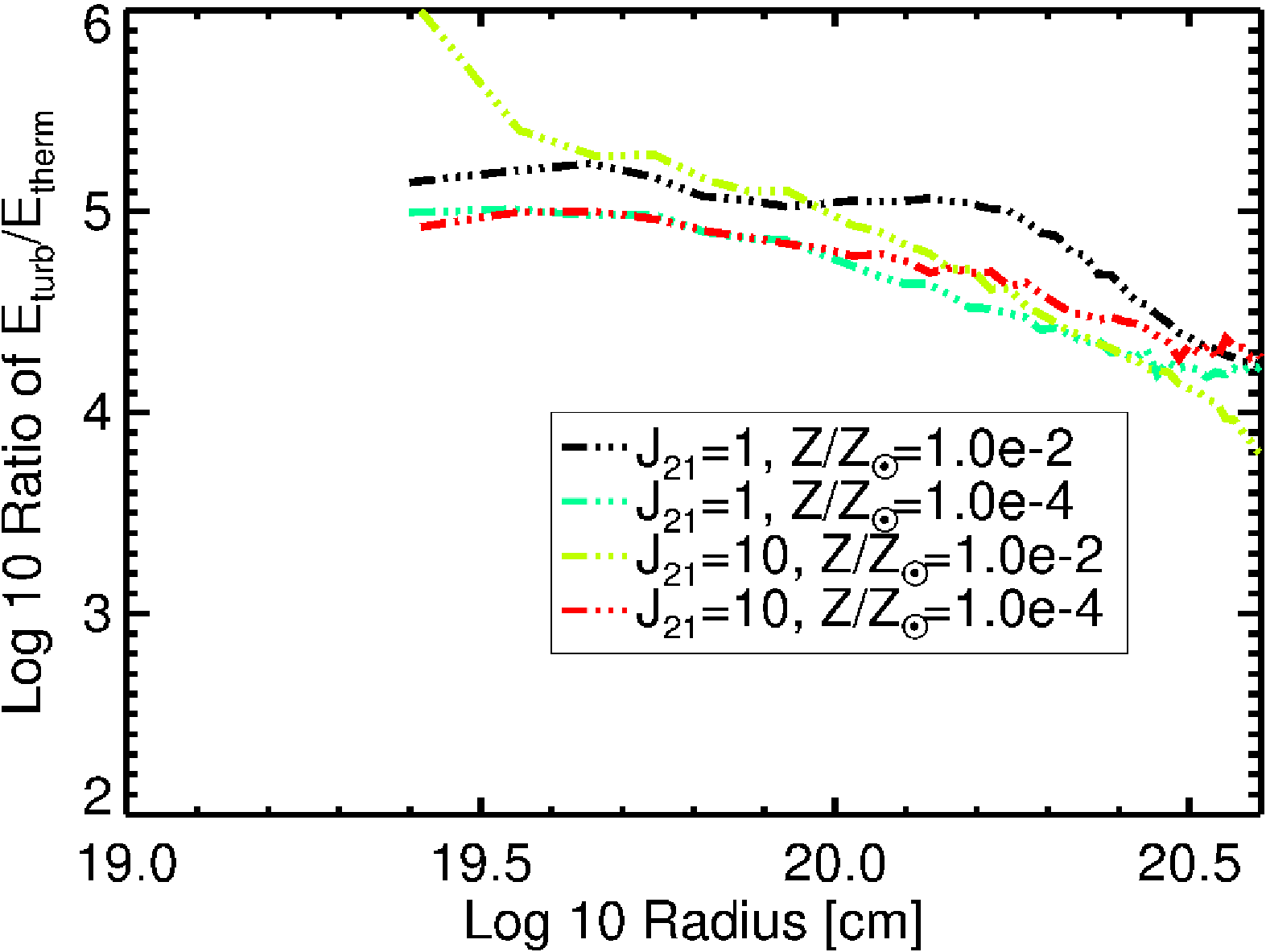}
 \end{minipage} &
\begin{minipage}{8cm}
\includegraphics[scale=0.28]{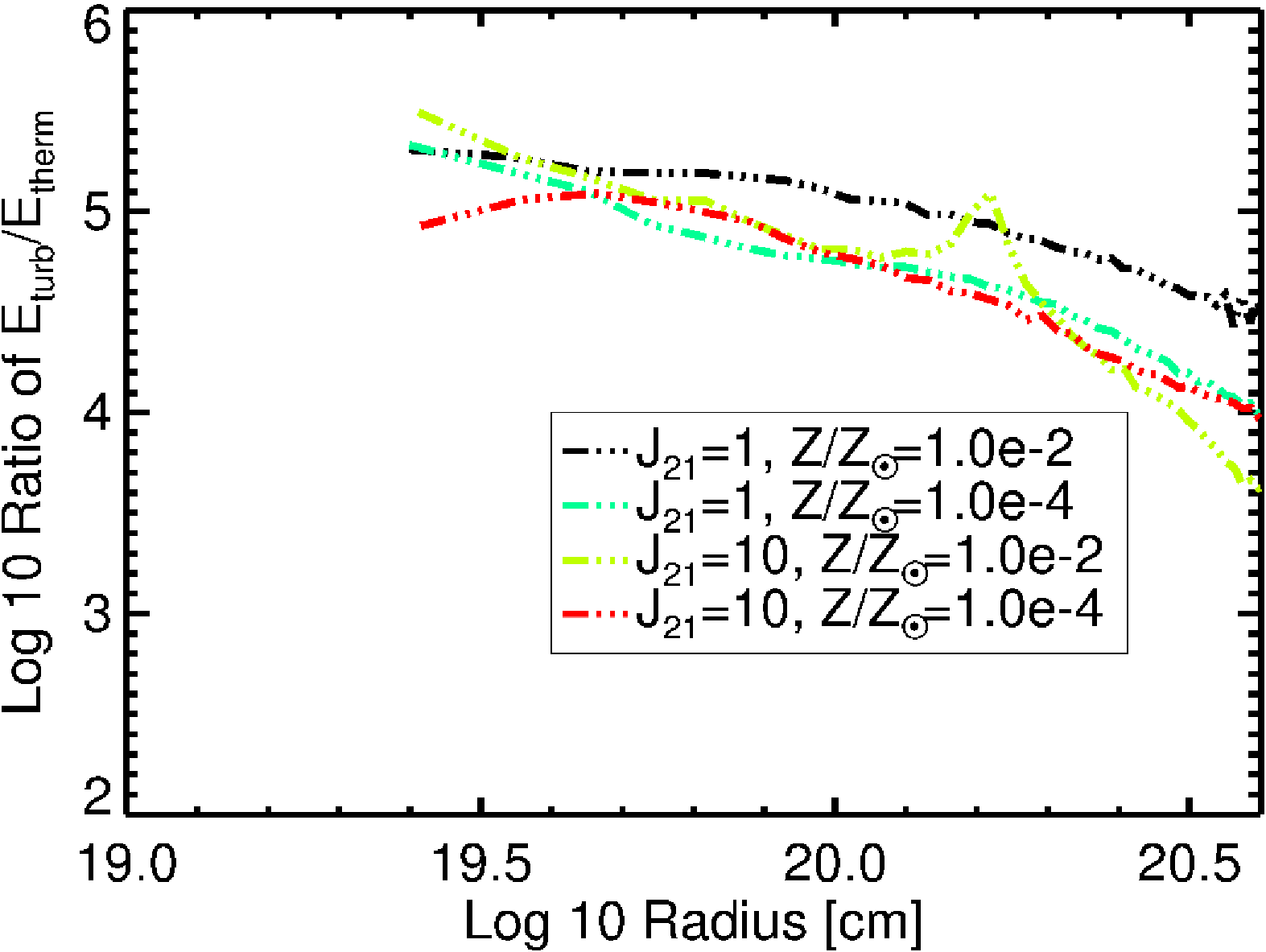}
\end{minipage} \\  \\

\begin{minipage}{8cm}
\includegraphics[scale=0.28]{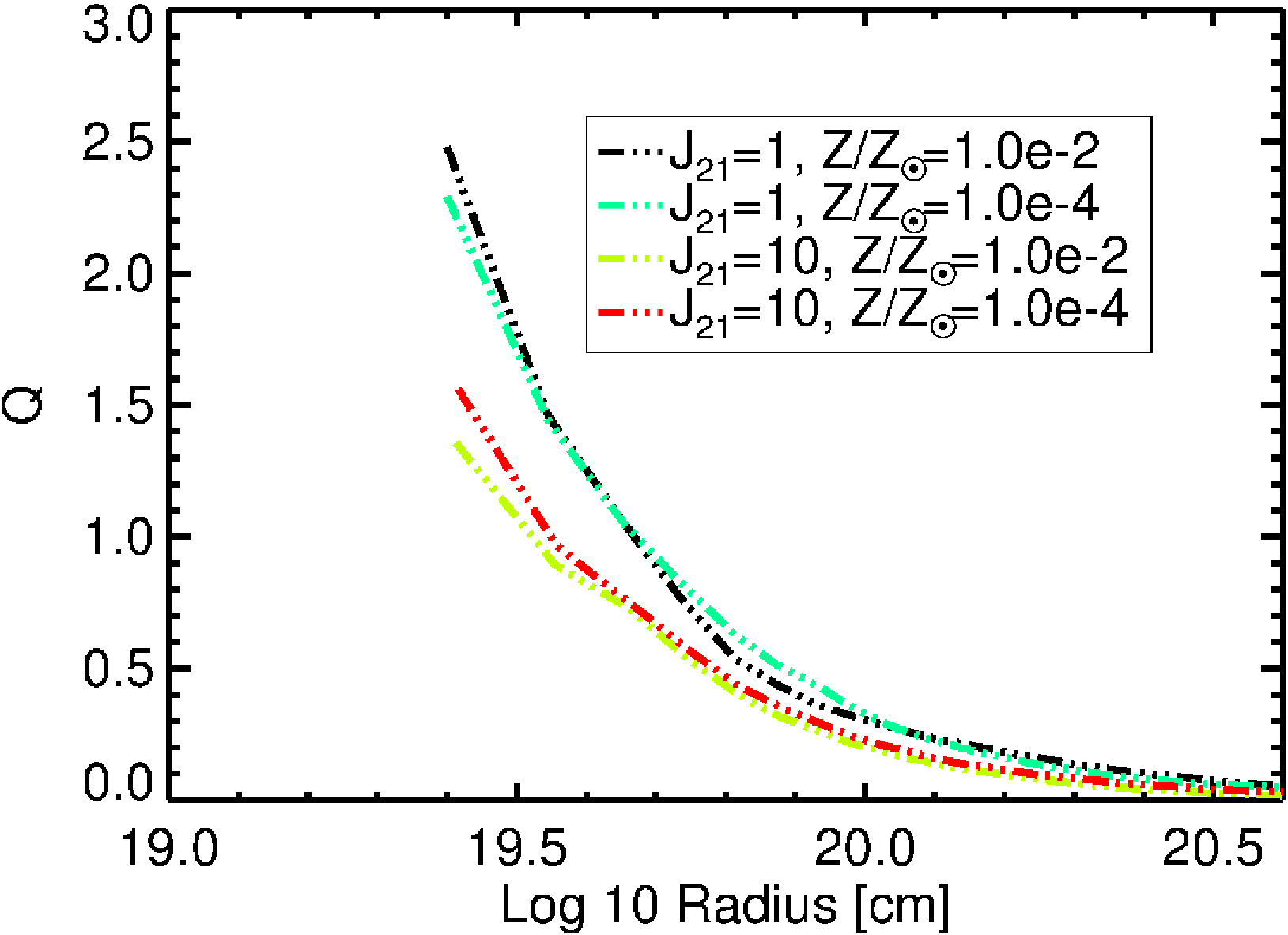}
\end{minipage} &

\begin{minipage}{8cm}
\includegraphics[scale=0.28]{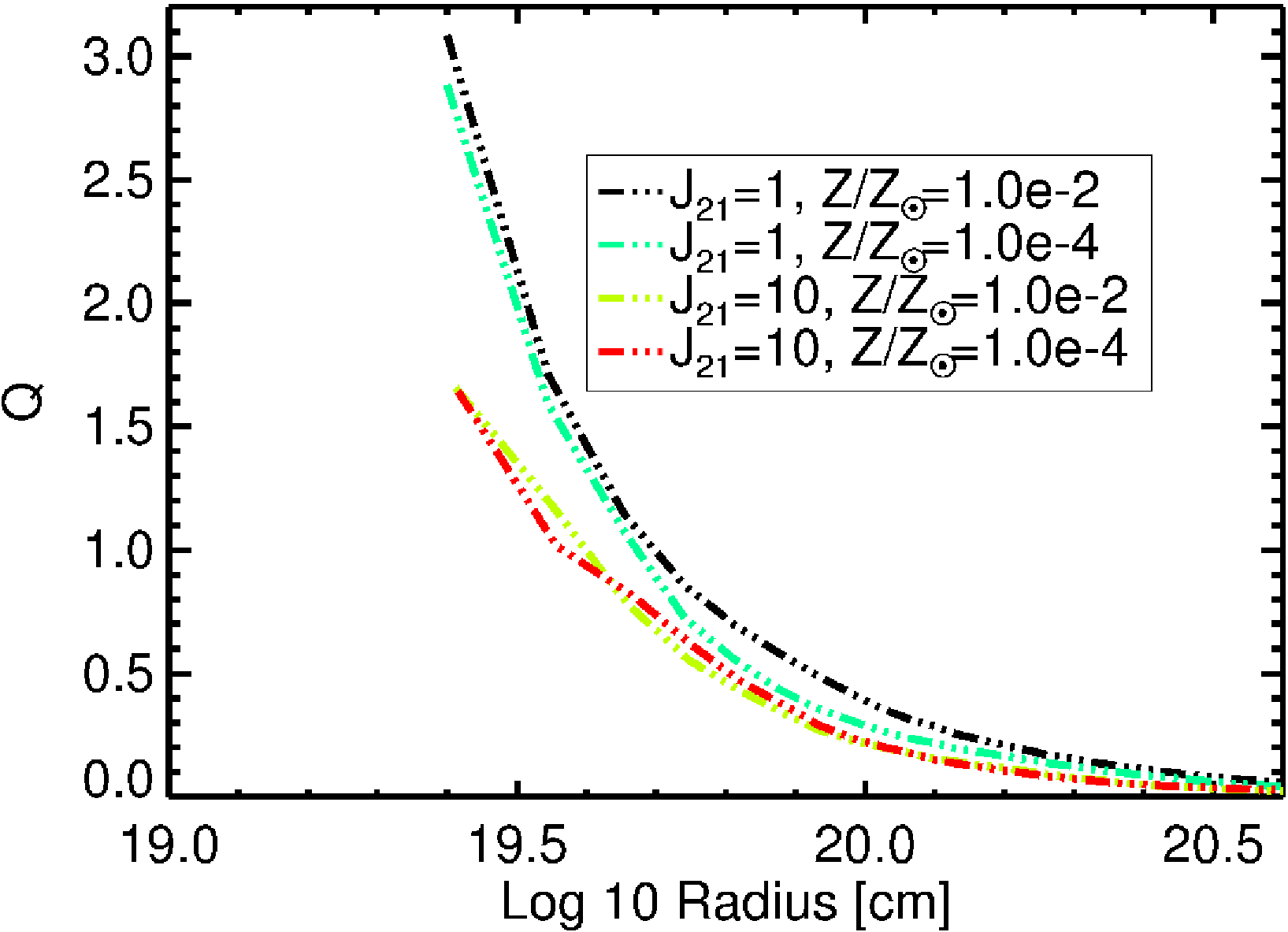}
\end{minipage}

\end{tabular}
\caption{Average radial profiles of disks for different background UV fields and dust-to-gas ratios at z=5.4. The left panels of this figure show radial profiles for disk 1 and the right panels show the radial profiles for disk 2. The top two panels show the ratio of rotational to orbital velocity for the disks. The middle panels show the ratio of turbulent to thermal energy for each disk. The Toomre parameter Q for each disk is depicted in the bottom panel. The escape fraction of ionizing radiation is 10\% for all panels.}
\label{figure3}

\end{figure*}

\begin{figure*}[htb!]
\centering
\begin{tabular}{c c}
\begin{minipage}{8cm}
% \hspace{0.27cm}
\includegraphics[scale=0.28]{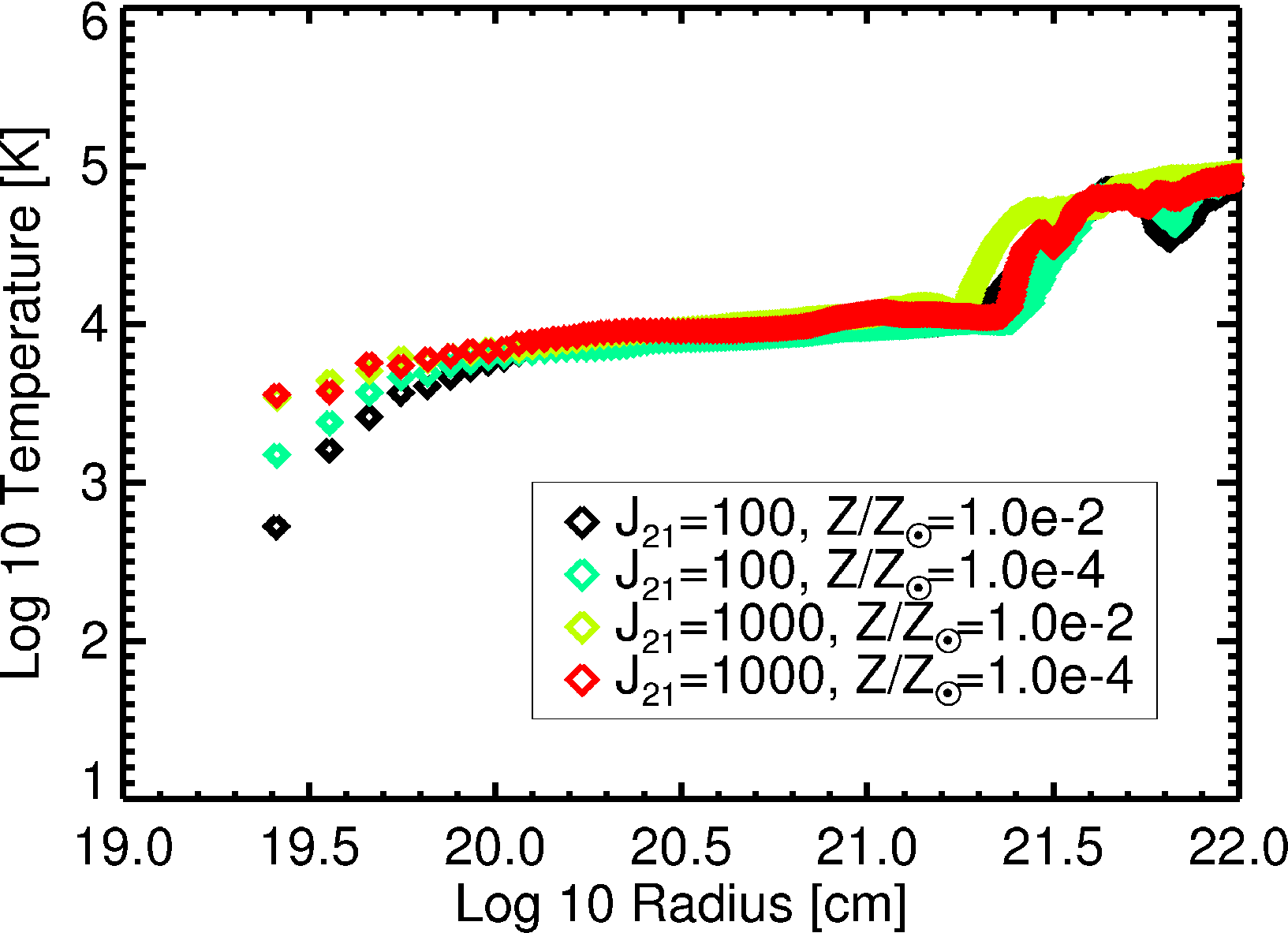}
\end{minipage} &
\begin{minipage}{8cm}
\includegraphics[scale=0.28]{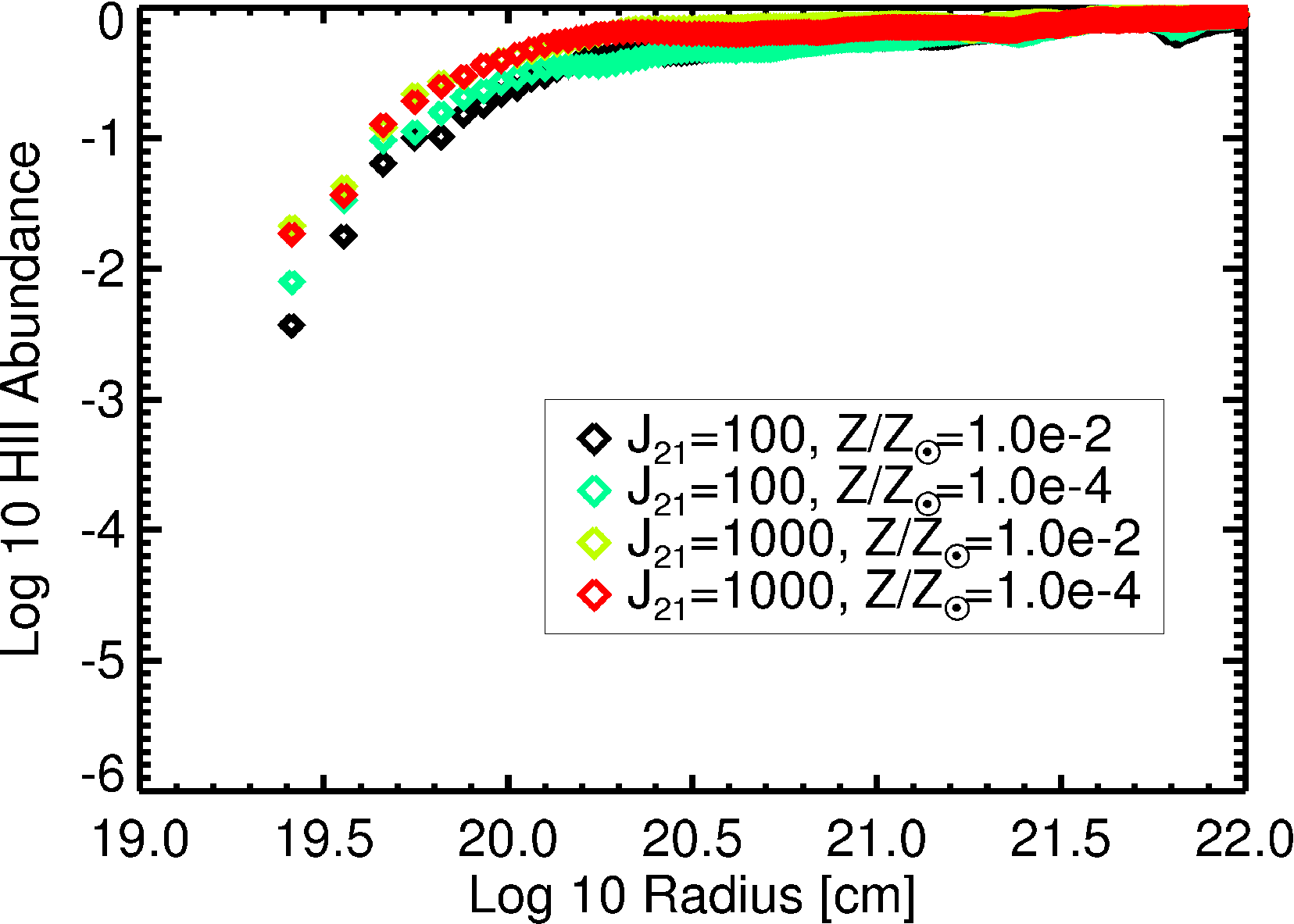}
\end{minipage} \\  \\

\begin{minipage}{8cm}
\includegraphics[scale=0.28]{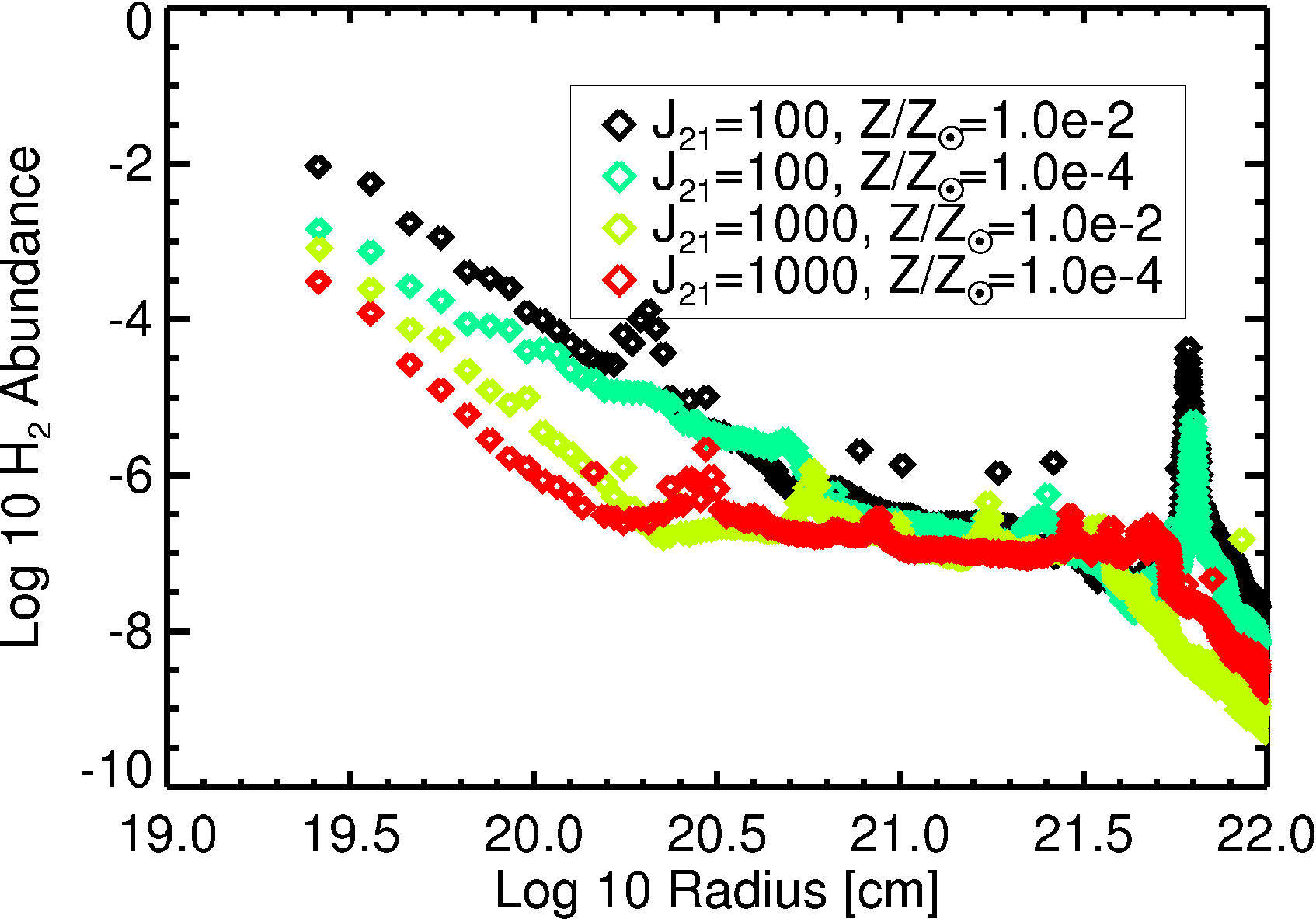}
\end{minipage} &

\begin{minipage}{8cm}
\includegraphics[scale=0.28]{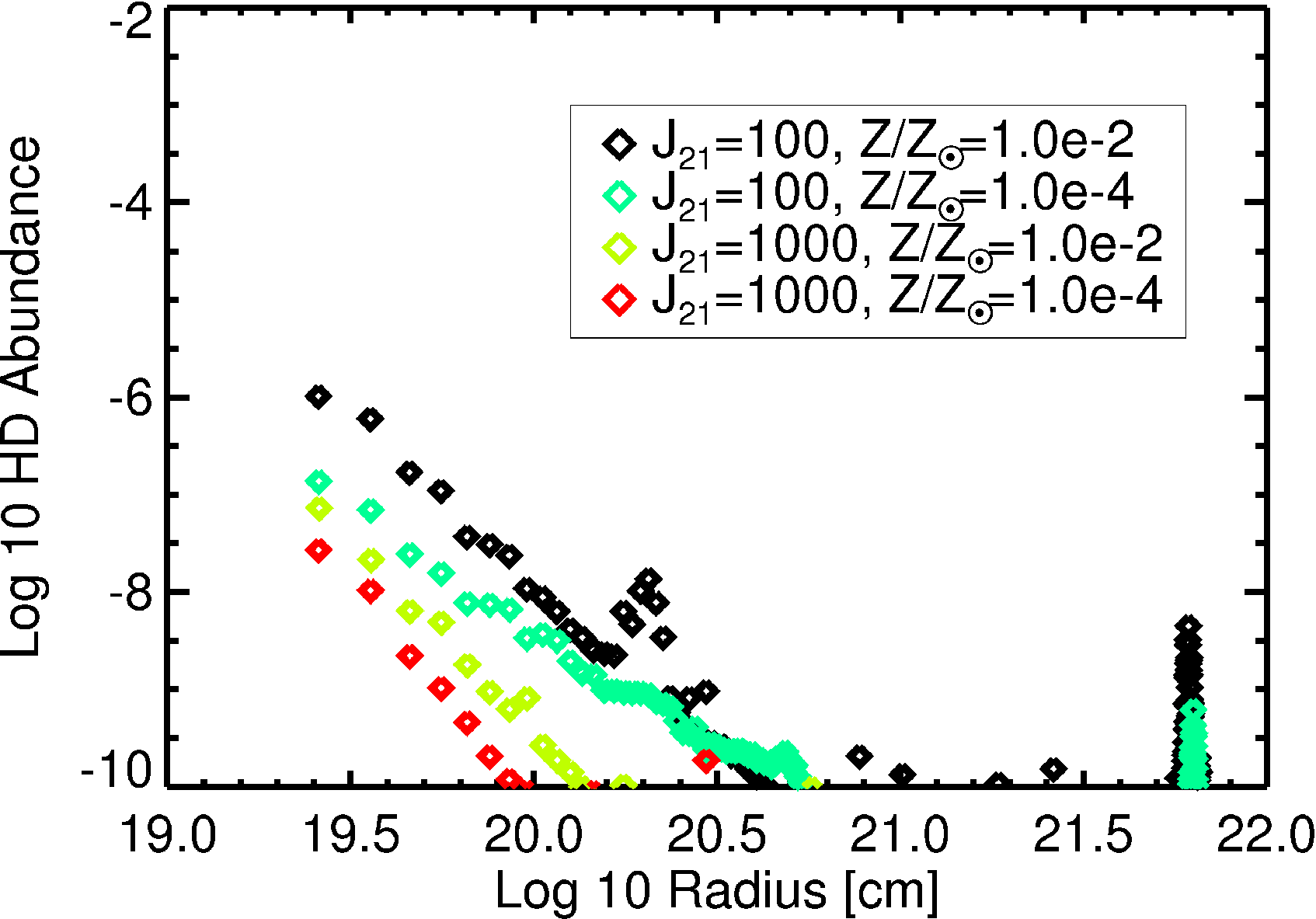}
\end{minipage}

\end{tabular}
\caption{Average radial profiles for different internal UV field strengths and dust-to-gas ratios at z=5.4. The upper left panel of this figure shows the temperature radial profile of the halo. The HII abundance radial profile for the halo is depicted in the upper right panel. The lower left panel shows the $\rm H_{2}$ abundance. The HD radial profile of the halo is depicted in the lower right panel. The escape fraction of ionizing radiation is 10\%.}
\label{figure4}
\end{figure*}

\begin{figure*}[htb!]
\centering
\begin{tabular}{c c}
\begin{minipage}{8cm}
% \hspace{0.27cm}
\includegraphics[scale=0.28]{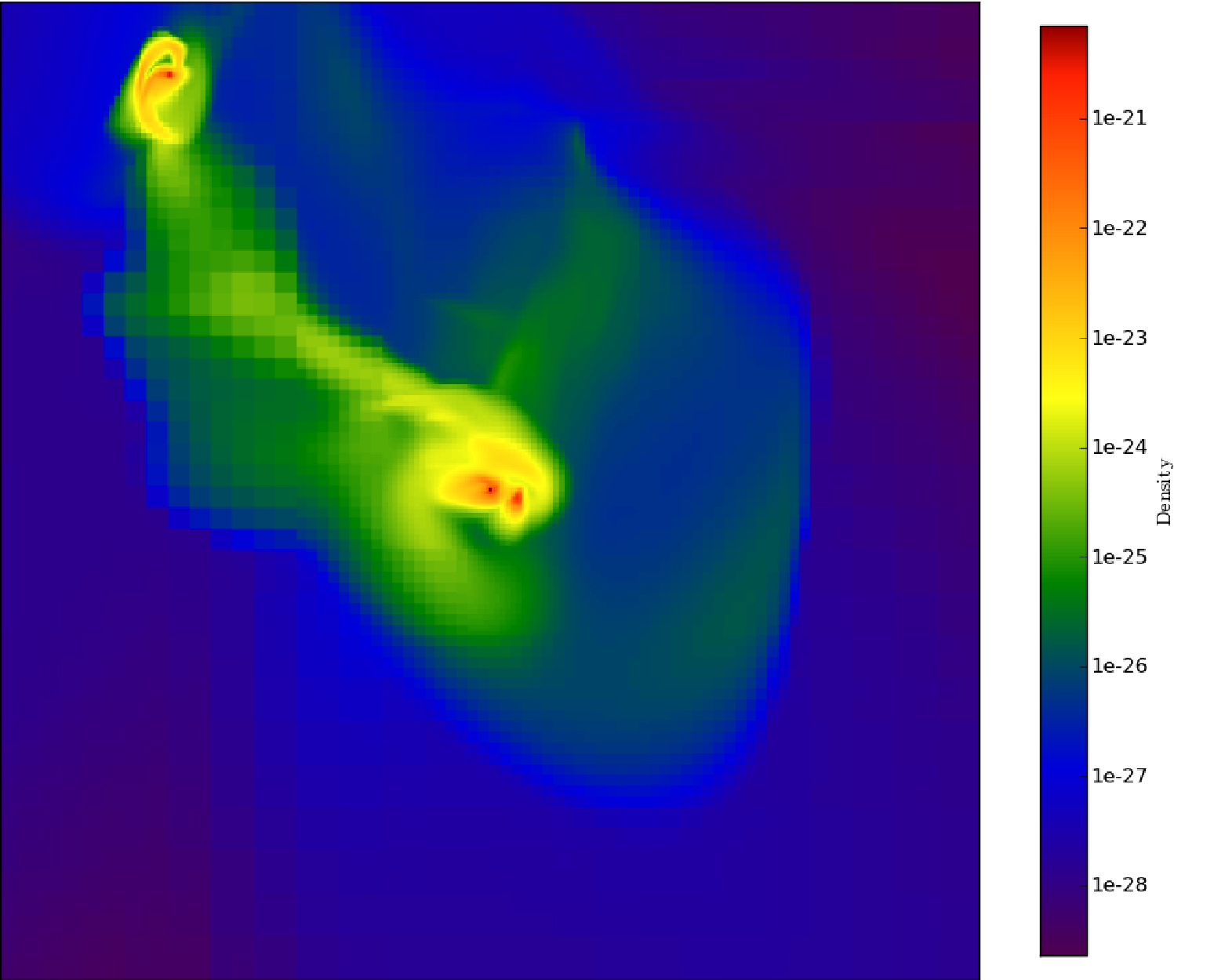}
\end{minipage} &
\begin{minipage}{8cm}
\includegraphics[scale=0.28]{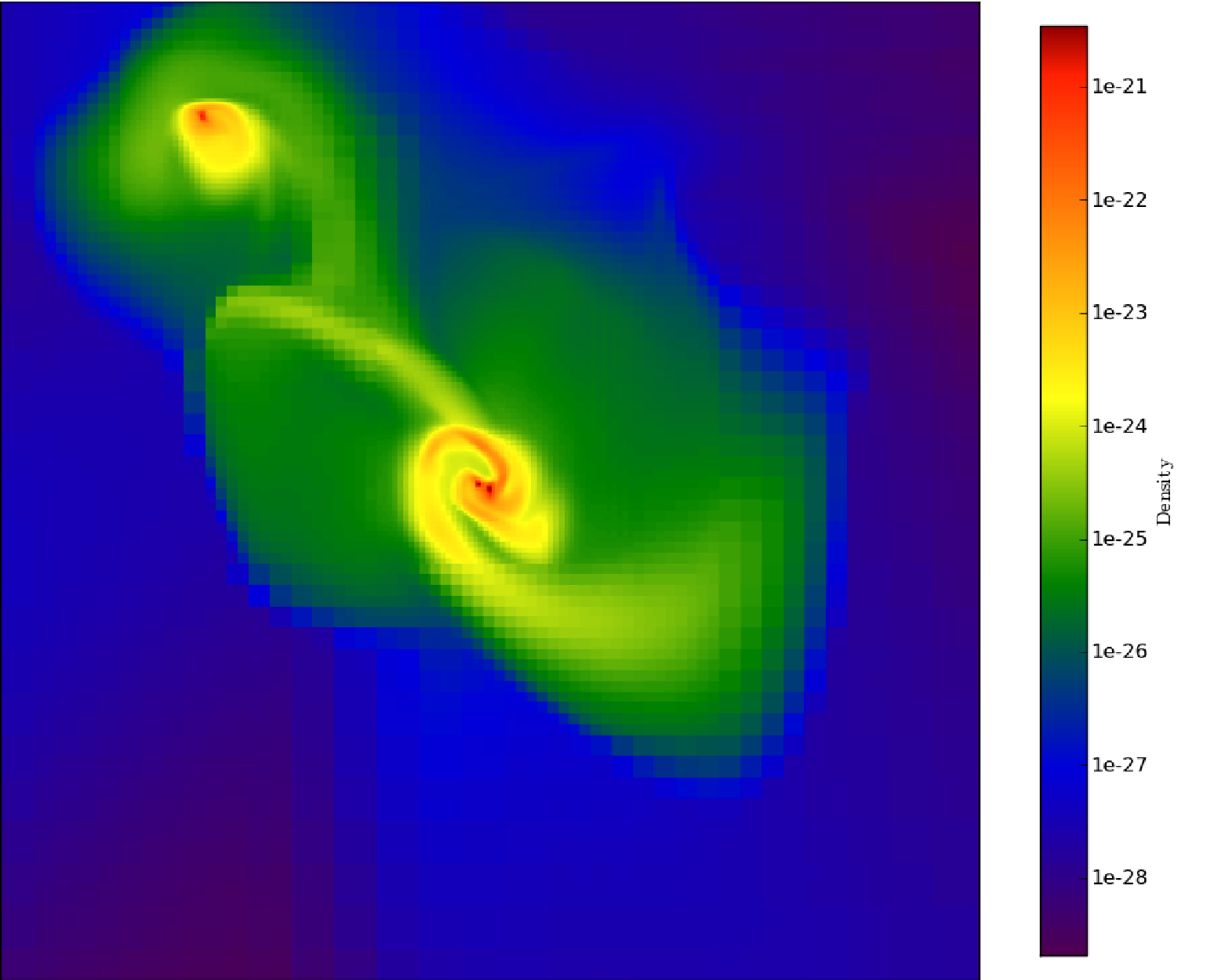}
\end{minipage} \\  \\

\begin{minipage}{8cm}
\includegraphics[scale=0.28]{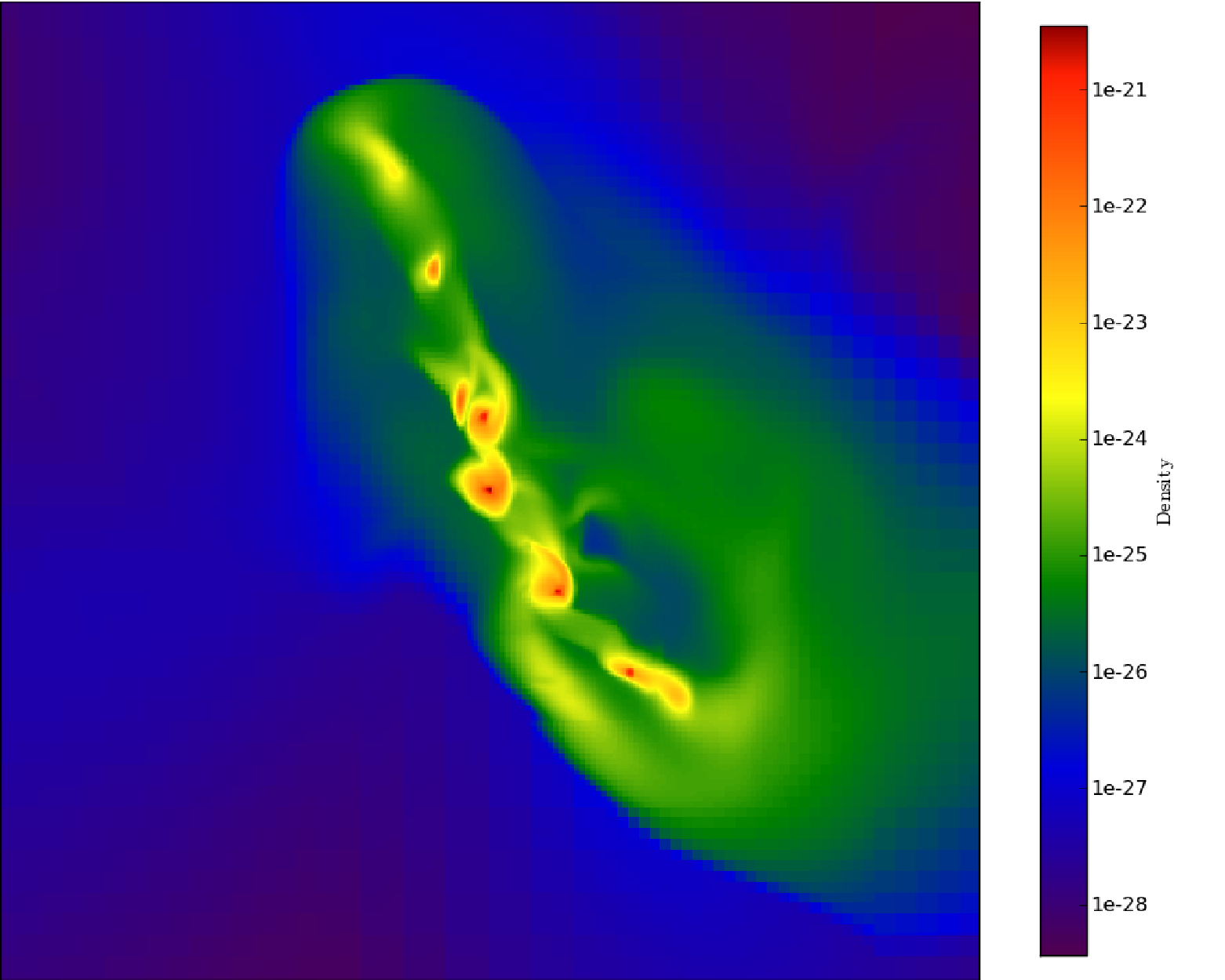}
\end{minipage} &

\begin{minipage}{8cm}
\includegraphics[scale=0.28]{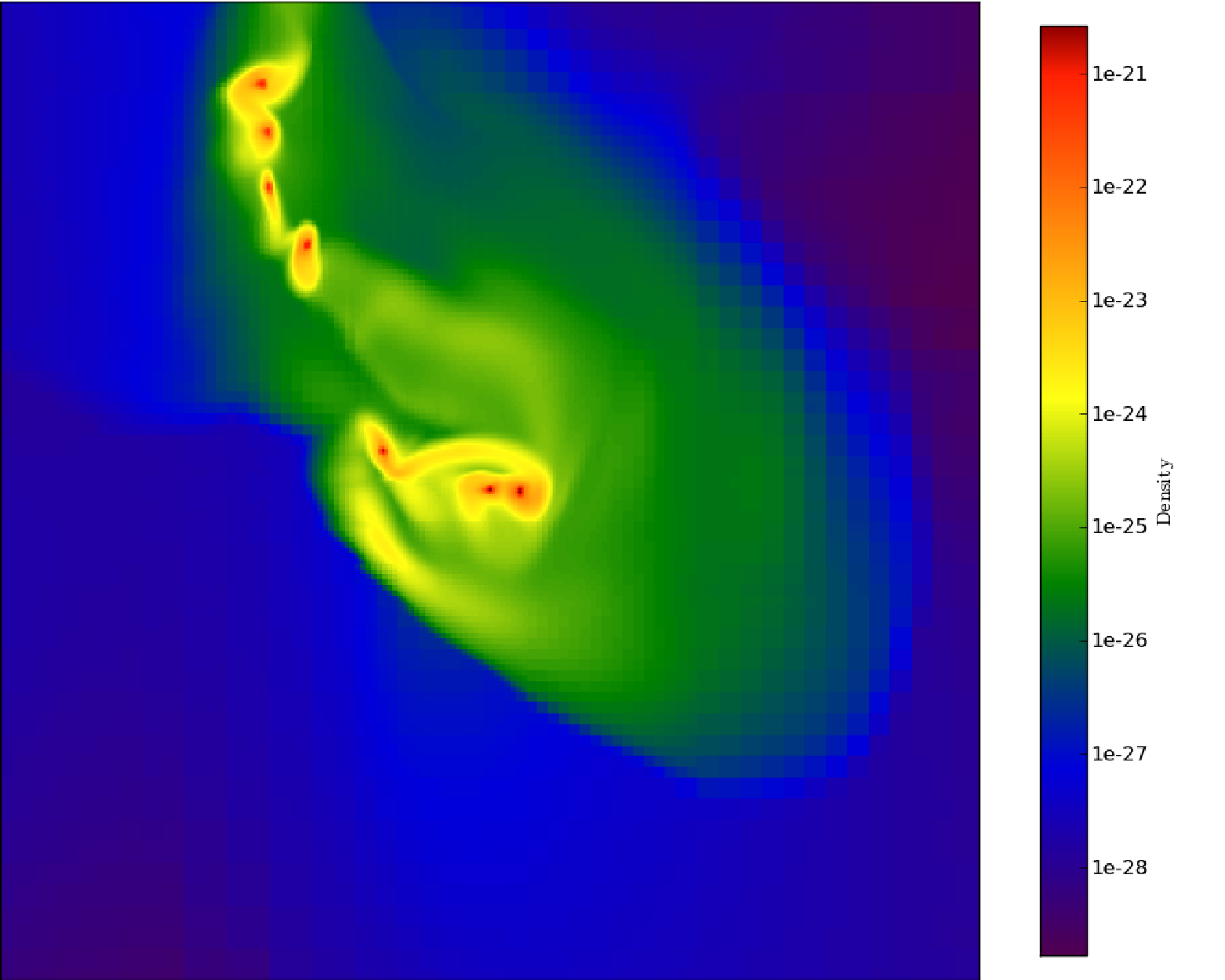}
\end{minipage} 
\end{tabular}
\caption{Density weighted projections of gas density for various internal UV field strengths and dust-to-gas ratios at redshift 5.4. The top left panel show the density for model E. The density structure for model F is depicted in the upper right panel. The left panel shows the density structure for model G. The state of a halo for model H is shown in the right panel. The escape fraction of ionizing radiation is 10\% for all panels. The values of density shown in the colorbar are in comoving units [$\rm g/cm^{3}$].}
\label{figuredd}
\end{figure*}

\begin{figure*}[htb!]
\centering
\begin{tabular}{c c}
\begin{minipage}{8cm}
\includegraphics[scale=0.28]{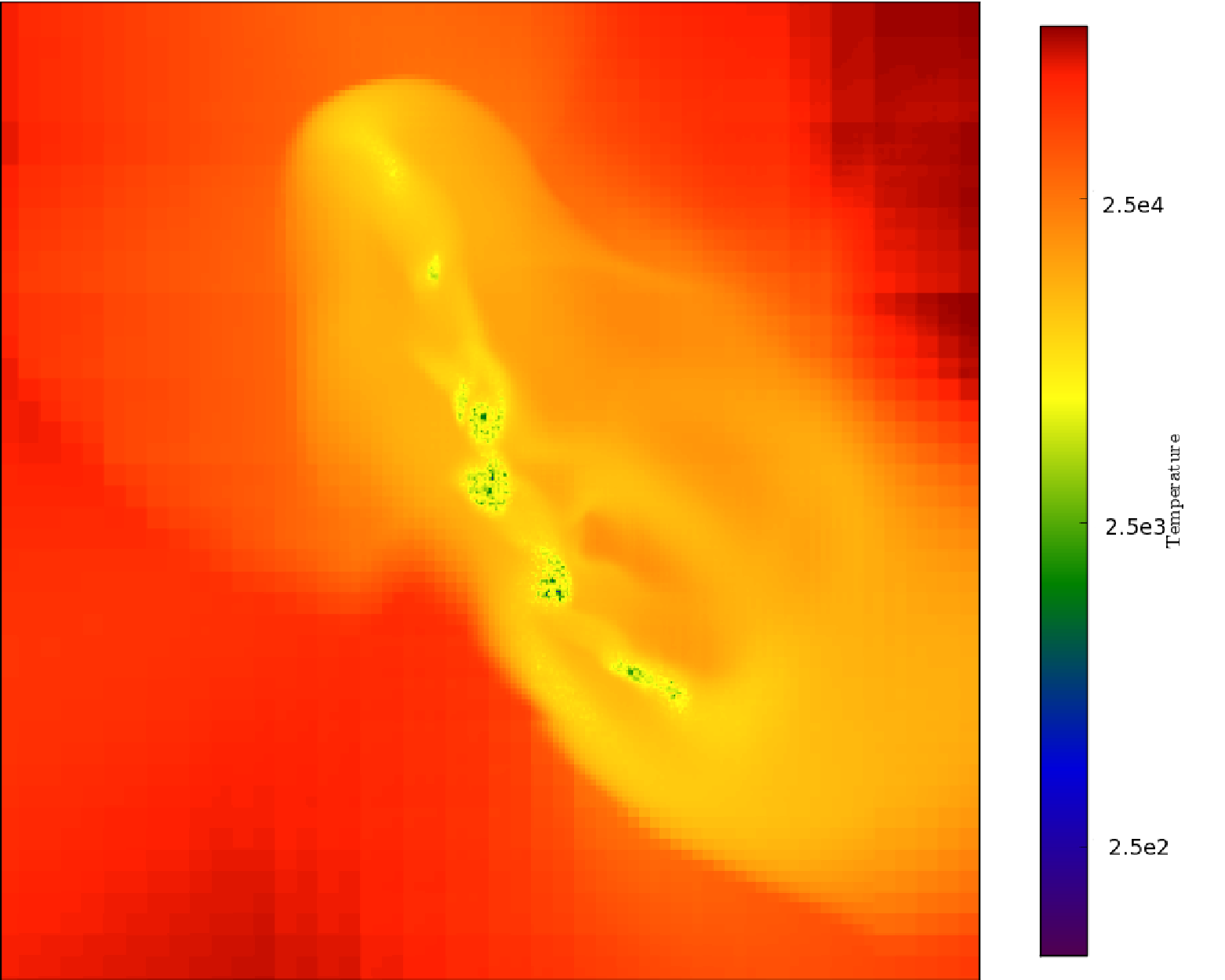}
\end{minipage} &

\begin{minipage}{8cm}
\includegraphics[scale=0.28]{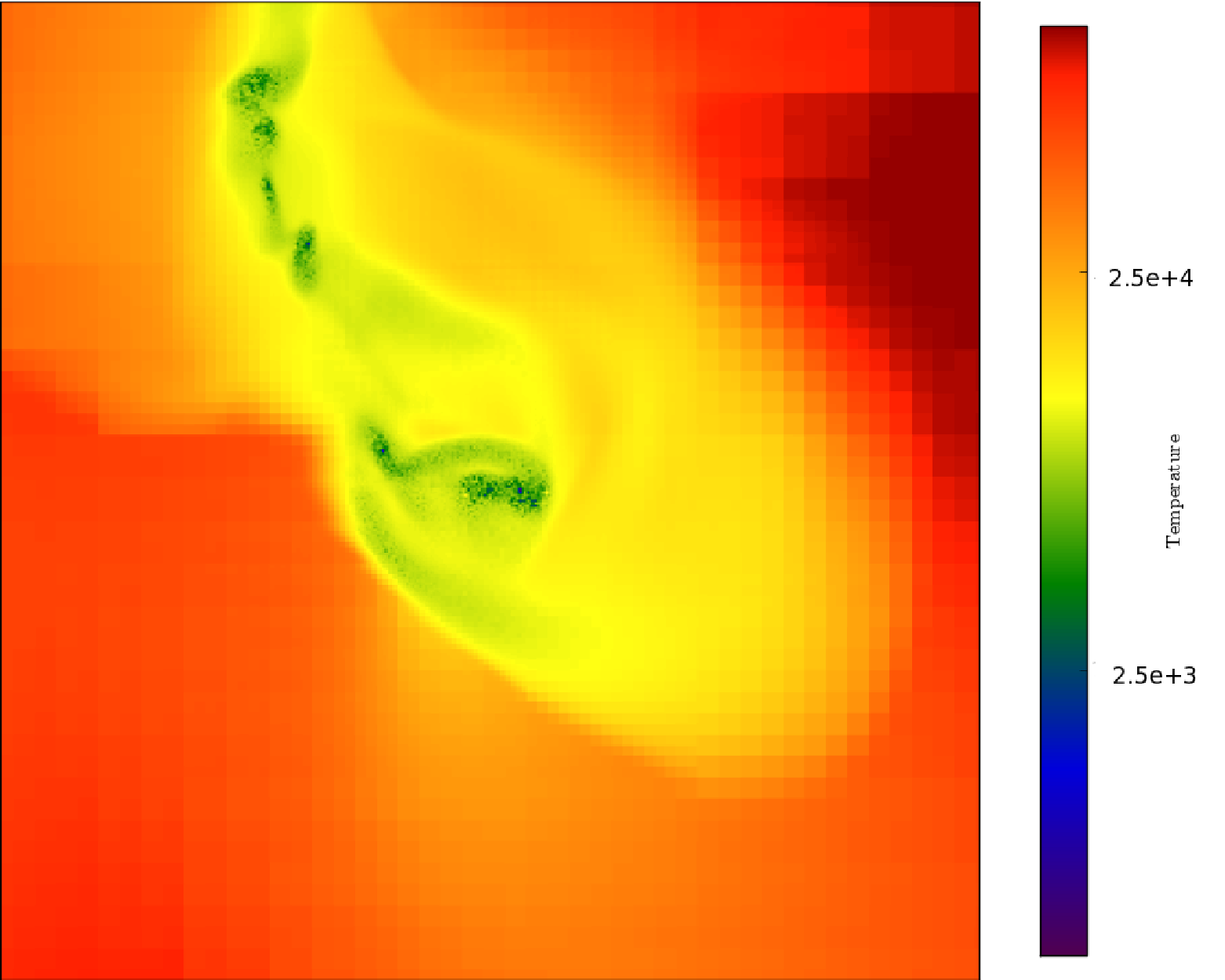}
\end{minipage}
\end{tabular}
\caption{Density weighted projections of temperature for different internal UV field strengths and dust-to-gas ratios at redshift 5.4. The left panel shows the temperature for model G. The temperature for model H is depicted in the right panel. The panels show temperatures corresponding to the density panels of figure \ref{figuredd}. This shows that thermal instability leads to the formation of gas clumps at these temperatures. The values of temperature are shown in the colorbar.}
\label{figured2}
\end{figure*}

\begin{figure*}[htb!]
\centering
\begin{tabular}{c c}
\begin{minipage}{8cm}
\includegraphics[scale=0.28]{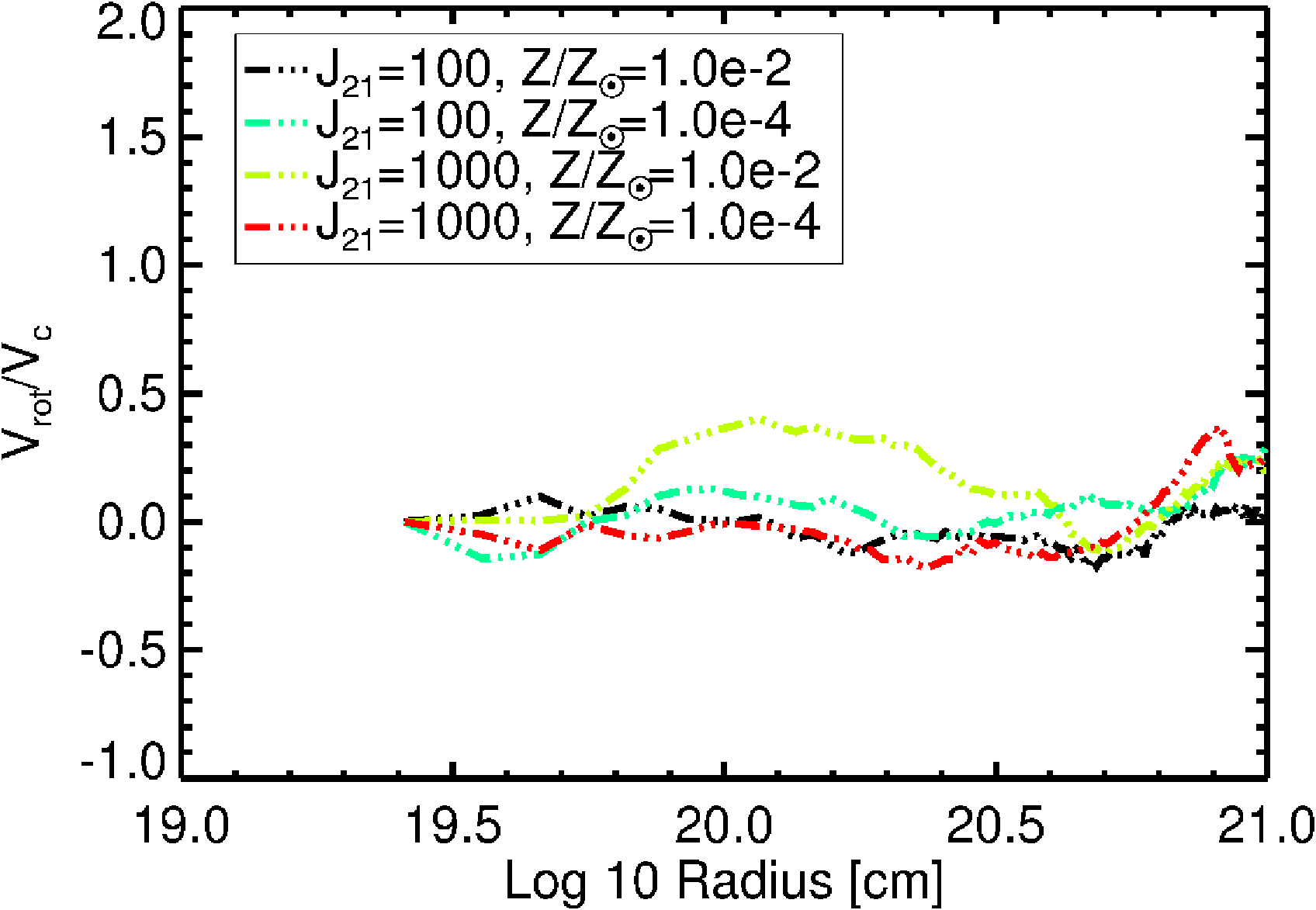}
\end{minipage} &
\begin{minipage}{8cm}
\includegraphics[scale=0.28]{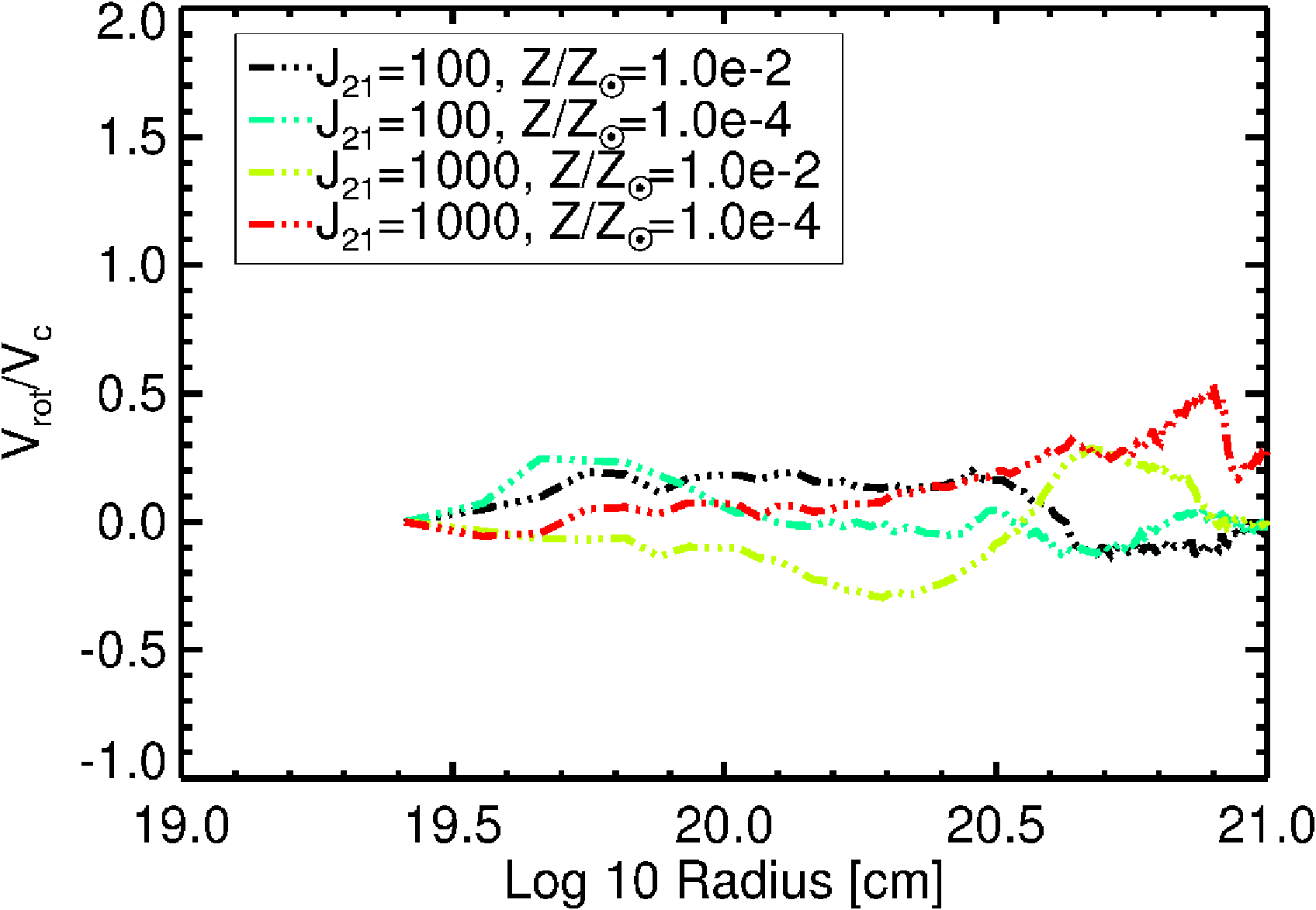}
\end{minipage} \\ \\
\begin{minipage}{8cm}
\includegraphics[scale=0.28]{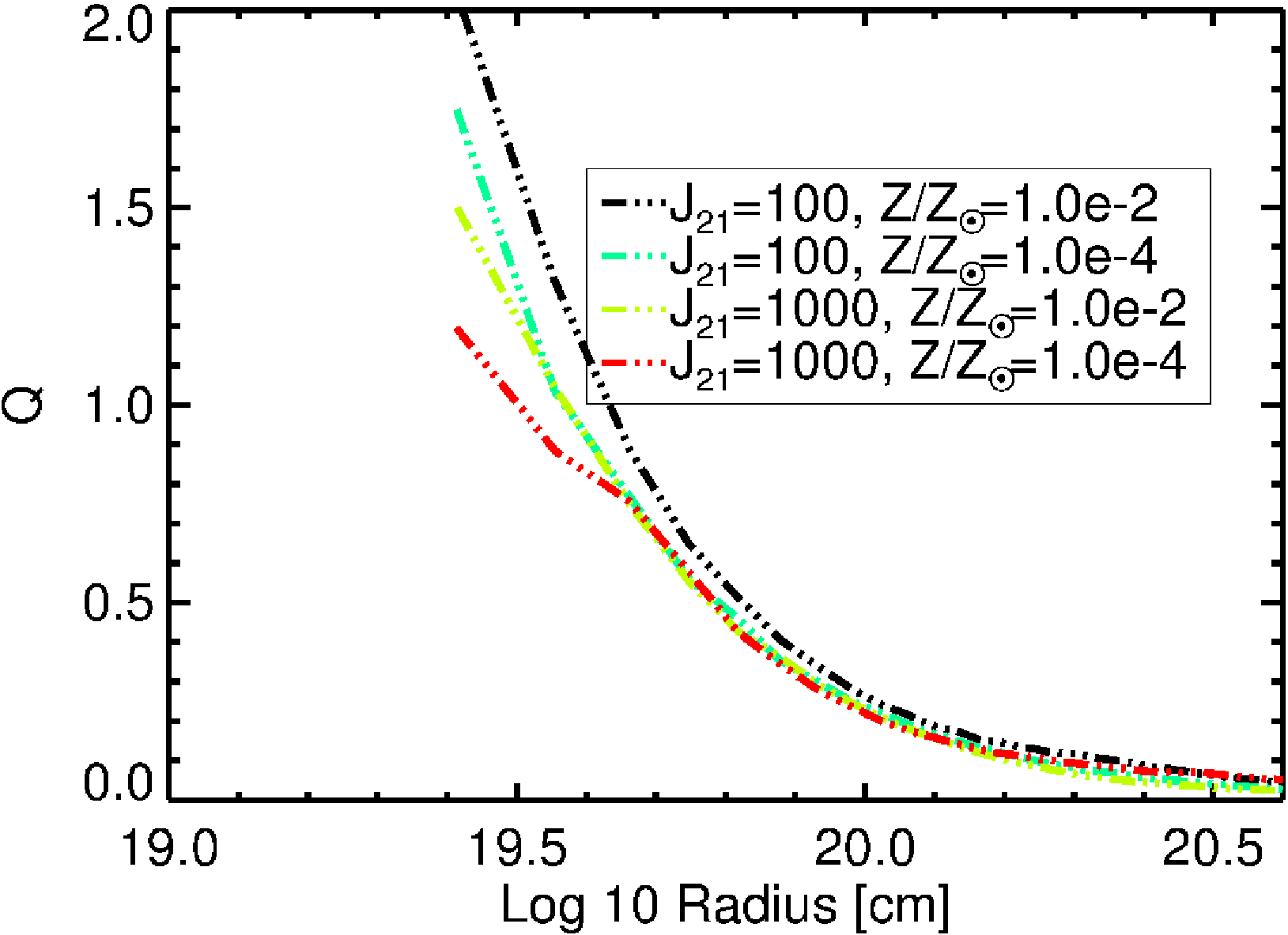}
\end{minipage} &
\begin{minipage}{8cm}
\includegraphics[scale=0.28]{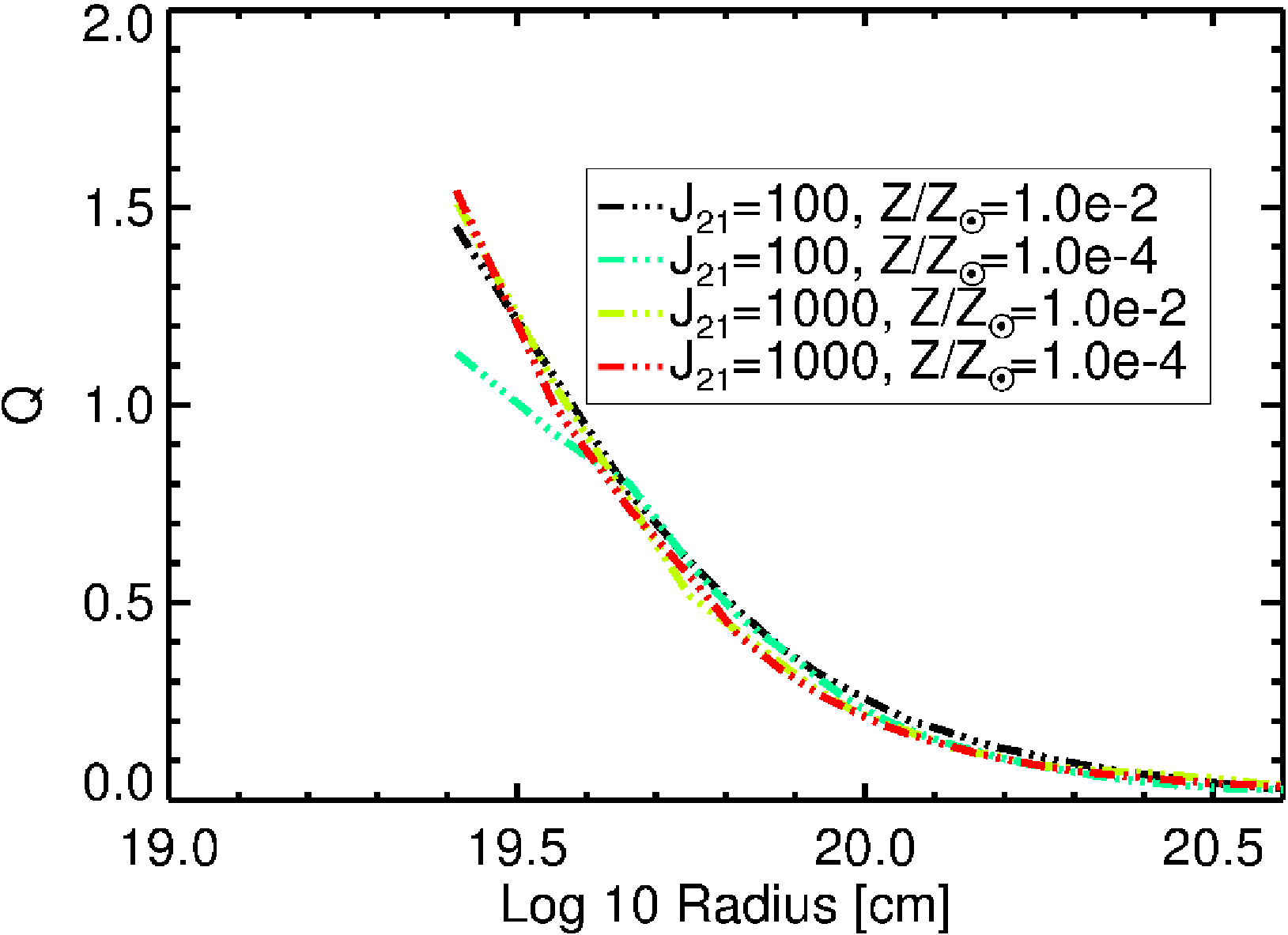}
\end{minipage}
\end{tabular}
\caption{Average radial profiles of the disks for different strengths of internal UV fields and dust-to-gas ratios at z=5.4. The left panel of this figure shows radial profiles for disk 1 and right panels shows the radial profiles for disk 2. The top two panels show the ratio of rotational to orbital velocity for the disks. The Toomre parameter Q for each disk is depicted in the bottom panels. The value of an escape fraction for ionizing radiation is 10\% for both panels.}
\label{figuret}
\end{figure*}

\begin{table*}[htb]

\begin{center}
\caption{Estimates of dust continuum emission for different UV radiation fields}
\begin{tabular}{cccccc}
\hline
\hline

Model	& Background UV flux			& Internal flux                    & Dust-to-gas ratio	                 & Dust temperature	\\

 & $\rm J_{21} [erg/cm^{2}/s/Hz/sr]$	&$\rm J_{21} [erg/cm^{2}/s/Hz/sr]$   & [$\rm Z/Z_{\odot}$]	                   & [$K$]	          \\ 
\hline
  A	  & 1				& 0	              & $1.0\times 10^{-2}$                        & 18.4                               \\		
  B	  & 1				& 0	  		& $1.0\times 10^{-4}$		           & 18.63		              \\
  C	  & 10				& 0	  		& $1.0\times 10^{-2}$	                   & 18.6		                \\	
  D	  & 10				& 0	  		 & $1.0\times 10^{-4}$	                   & 18.7		               \\
  E      & 1                           & 100     		   & $1.0 \times10^{-2}$                   & 22.4                               \\
  F      & 1                           & 100      		  & $1.0\times 10^{-4}$                    & 22.5                                  \\
  G	 & 1                           & 1000     		    & $1.0\times 10^{-2}$                  & 33.2                               \\
  H      & 1                           & 1000     		   & $1.0\times 10^{-4}$                    & 33.6                               \\

% 9       & 0.1                                   & 0.5                     & $10^{46}$                             & $5 \times 10^{-15}$    \\

\hline
\end{tabular}
\label{tab:sumary}
\end{center}

\end{table*}

\section{Results}

\subsection{Global dynamics}

We have performed a grid of simulations for different strengths of external and internal UV fluxes as well as for two different dust-to-gas ratios as shown in table \ref{tab:sumary}.
We start our simulations at redshift 90 with cosmological initial conditions. Density perturbations decouple from the Hubble flow and begin to collapse through gravitational instability at redshift 20. The gas falls into the dark matter potentials and gets shock heated. The virialization processes partly transform the gravitational binding energy into kinetic energy of the gas and dark matter. Part of the gravitational potential energy goes into thermal energy of the gas and is radiated away. We also see the merging of smaller clumps into bigger ones. The occurrence of a major merger is shown in figure \ref{figured1}. This process of merging helps in the transfer of angular momentum, which is necessary for the gas to settle in the gravitational potential \citep{2010Natur.466.1082M}. We switch on the radiative backgrounds at $\rm z=8$, where we expect reionization to be already in its final phase. The temperature of the gas is increased above $\rm 10^{4}~K$ due to photoheating, depending on the strength of the radiation field. At these temperatures, the recombination cooling kicks in and the gas begins to recombine. The gas in the filaments is relatively denser, becomes self-shielded and falls into the center of the galaxy through cold streams \citep{2000ApJ...537L...5H,2009MNRAS.400.1109D}. These cold streams have typical temperatures of $\rm 10^{4}~K$ and number densities on the order of $\rm 0.01-1~cm^{-3}$. The average density radial profiles for different strengths of the background UV fields are shown in figure \ref{figure1}. The variations in the density radial profiles are due to the substructure in the halo and vary with the strength of the radiation field and dust-to-gas ratio. Before discussing the disk morphologies and dynamics in more detail, we will explore the impact of dust and radiation backgrounds in section 4.2.
% Ionizing flux is produced during the epoch of reionization by the stellar sources at redshift about 8. As we turn on the ionizing radiation flux, it photoionizes the gas.
\subsection{Chemical and thermal evolution}

The average temperature radial profiles for two different dust-to-gas ratios and various strengths of a background UV radiation field are shown in the upper left panel of figure \ref{figure2}. It can be seen that gas is photoheated by a UV radiation flux in the envelope of a halo and consequently its temperature is raised to above $\rm 10^{4}~K$. It can also be seen that the maximum temperature is $\rm 10^{5}~K$ for ionizing flux ($\rm J_{21}\times f_{esc}$) of 1 in units of $\rm J_{21}$. The cooling due to helium lines becomes important at $\rm \geq 5\times10^{4}~K$ and thus keeps the gas temperature at $\rm 10^{5}~K$. Recombination cooling becomes effective above $\rm 10^{-24}~g/cm^{3}$ and cools the gas down to $\rm 10^{4}$ K. It is found that for model A, gas cools efficiently due to the enhanced fraction of $\rm H_{2}$ and HD molecules in the presence of dust grains. The $\rm H_{2}$ and HD cooling begins to become effective around $\rm 10^{-22}~g/cm^{3}$, where gas is cooled down to $\rm \sim$150 K. For model B, the temperature in the center of the halo is increased to approximately 250 K. Similarly, for models C and D (i.e., $\rm J_{21}=10$) the temperature is higher in the center of a halo due to the photodissociation of $\rm H_{2}$ and HD molecules.

The temperature evolution is significantly different for higher and lower dust-to-gas ratios. The ionization degree of the gas is shown in the top right panel of figure \ref{figure2}. The ionization fraction of HII is higher in the surroundings of the halo, depending on the intensity of ionizing radiation, and goes down towards the center of the halo. The overall degree of ionization is higher for a stronger radiation field and lower for a weaker radiation field. At radii $\rm < 10^{19.7}~cm$, the ionization fraction decreases sharply as gas becomes neutral and is efficiently shielded from the ionizing radiation flux. The results are in agreement with our previous study \citep{2011A&A...532A..66L}. 

The bottom left panel of figure \ref{figure2} depicts the $\rm H_{2}$ radial profiles. It is found that the molecular hydrogen abundance is very low in the envelope of the halo as gas is highly ionized. There is a sharp rise in $\rm H_{2}$ abundance at radii between $\rm 10^{21.5}$ cm and $\rm 10^{22}$~cm. This is due to the substructure in the halo where gas self-shields against the background UV radiation field and molecular hydrogen survives. The $\rm H_{2}$ abundance sharply increases towards the center of the halo as its photodissociation becomes ineffective due to further self-shielding. For a weaker radiation field and higher dust-to-gas ratio (i.e., model A), the $\rm H_{2}$ fraction is higher compared to the stronger radiation field and lower dust-to-gas ratio cases. The $\rm H_{2}$ abundance is significantly boosted in the core of the halo due to the formation of $\rm H_{2}$ molecules on dust grains at high gas densities. This increase in $\rm H_{2}$ abundance in the presence of dust grains is consistent with earlier studies of \cite{2009A&A...496..365C,2004ApJ...611...40C}. Similar behavior is found for HD molecules, as shown in the bottom right panel of figure \ref{figure2}. The HD abundance is lower for a stronger UV background flux and higher for a weaker flux. A maximum HD abundance of $\rm 10^{-4.5}$ is found for model A. The HD abundance remains lower than the critical value (i.e., $\rm 10^{-6}$) for $\rm J_{21}\geq 0.1$ in agreement with previous studies \citep{2011MNRAS.412.2603W,2011A&A...532A..66L}.

\subsection{Disk formation and morphology} 

The properties of simulations for different strengths of background UV flux and dust-to-gas ratios at redshift 5.4 are illustrated in figure \ref{figured}. It is found that binary disks are formed for model A, as shown in the top left panel of figure \ref{figured}, in different subhalos. We note that the upper disk is formed first and the second disk appears a few megayears after the formation of first. Each disk has spiral arms and gas masses of $\rm 4 \times 10^{7}~M_{\odot}$ and $\rm 2 \times 10^{7}~M_{\odot}$, respectively, and sizes of $\rm \sim$ 1 kpc. We see indications for the formation of an additional disk during later stages of the evolution. For the same radiation field, but with lower dust-to-gas ratio (i.e., model B), the halo fragments again and forms binary disks. The morphology of the halo is shown in the top right panel of figure \ref{figured} and is very different from the previous case. The temperature of the halo is higher due to the lower dust-to-gas ratio. The gas masses of the lower and upper disks are $\rm 2.3 \times 10^{7}~M_{\odot}$  and $\rm 1.4 \times 10^{7}~M_{\odot}$, respectively. We found that the lower disk is formed first (opposite to the previous case) and the second disk is formed after 6 megayears. The bottom left panel of figure \ref{figured} shows the structure of a halo for model C. It can be seen that again binary disks are formed. The formation process is different, the top disk is formed first while the second disk is formed through the merging of smaller clumps. Moreover, the formation of the bottom disk is delayed compared to the lower radiation background UV field case and is formed 19 megayears after the formation of the first disk. The overall collapse of the halo is also delayed due to the stronger radiation field. The gas masses of the lower and upper disks are $\rm 2.0 \times 10^{7}~M_{\odot}$ and $\rm 3\times 10^{7}~M_{\odot}$, respectively. The bottom right panel of figure \ref{figured} shows the state of the halo for model D. Similar to the other cases, two disks are formed. The upper disk is formed first and the second disk is formed  after 20 megayears. This delay is due to the enhanced strength of the radiation field and lower dust-to-gas ratio. Gas masses of the lower and upper disks are $\rm 2.2 \times 10^{7}~M_{\odot}$, $\rm 1.7\times 10^{7}~M_{\odot}$, respectively. The binary disk formation in our simulations is due to the presence of subhalos after a major merger. Employing different initial conditions may result in single or multiple disks.

\subsection{Disk dynamics and stability}

We have computed the circular velocity ($\rm v_{c}=\sqrt{GM/R}$, M is the disk mass) and rotational velocity ($\rm v_{rot}=L/R$, where L is the specific angular momentum of the disk) for each disk. The ratio of rotational to circular velocity for disk 1 (the massive disk in each case) is shown in the top left panel of figure \ref{figure3}, while for disk 2 (the less massive one in each case) is shown in the top right panel of figure \ref{figure3}. The rotational velocity remains lower than the circular velocity which means that the collapse is never halted by rotational support. In the very central region, there is a central massive clump with no spiral arm structure, and with very little rotational support. This is because the turbulent velocities are much larger than the rotational velocity, leading to a spherical rather than a disk-type structure. At larger scales $\rm >10^{19.6}~cm$ , the rotational velocity is increased and we see a disk with spiral arms. We found that both disks are co-rotating, independent of metallicity. The sign change in the rotational velocity is a result of the two spiral arms, as the arms between the disks then move in opposite directions. We found that disk 2 in each case has higher rotational support than disk 1. We also computed the turbulent energy for the system and compared it with the thermal energy. The ratio of turbulent to thermal energy is plotted for both disks in the middle panel of figure \ref{figure3}. It is found that turbulence is highly supersonic and turbulent pressure is more important than the thermal pressure. This is due to the high virial temperature, which is reflected in the kinetic energy of the gas. Our results are similar to \cite{2011ApJ...737...63A}. To investigate the stability of each disk, we computed the Toomre instability parameter Q, given by 
% \vspace{-0.327 cm}
\begin{equation}
\rm Q =  {c_{g}\Omega \over \pi G \Sigma } ,
% \sim 100 \left({T \over 100~K }\right)^{1/2}  \left( {\Sigma_{g}  \over 10^{3} cm^{-2}} \right)^{-1} \left( {R \over 1 AU}\right)^{-3/2} 
\label{Toomre}
\end{equation}
where  $\rm \Omega $ is the Keplerian orbital frequency, $\rm \Sigma_{g}$ is the surface density of the disk, $\rm c_{g}= \sqrt{c_{s}^{2} + v_{rms}^{2}}$, $\rm c_{s}$ is the isothermal sound speed and $\rm v_{rms}$ is the root mean square velocity $\rm v_{rms}=\sqrt{3/2 v_{tang}}$  and G is the gravitational constant. The bottom left panel of figure \ref{figure3} shows the spatial evolution of the Q parameter for disk 1. It is found that only for $\rm R < 10^{19.6}$ cm the value of Q is larger than one and the disk is stable. The sharp increase towards the center of the disks is due to the increase in epicyclic frequency. The Toomre instability parameter Q for disk 2 is shown in the bottom right panel of figure \ref{figure3}. The value of the Toomre parameter is not well defined on large scales (i.e., $> 10^{20}$ cm), as there is no well defined disk and the overall spatial evolution of mass is much larger than the Jeans mass. We also computed the specific angular momentum spatial evolution. We found that angular momentum increases with radius according to expectation. We have a specific angular momentum on the order of $\rm 10^{31}~cm^{2}/s$ and our results are again in agreement with \cite{2011ApJ...737...63A}. Our stability analysis shows that disks are stable and are supported by turbulent pressure. In spite of the global stability, the high turbulent Mach numbers considered will locally produce strong overdensities in shocks, and may lead to a star formation mode regulated by turbulence \citep{2004RvMP...76..125M,2007ApJ...654..304K,2011arXiv1104.5582B}. The final fate of the disks depends on the ambient conditions. They may merge with each other and lead to the formation of a protogalaxy. They may form stars if they become Toomre unstable.

% The temporal evolution of specific angular momentum and Toomre parameter are shown in figure \ref{figuret}.
\subsection{The implications of internal radiation fields}

We also performed simulations with a given strength (i.e., $\rm J_{21}=1$) of the background UV radiation field, but for an enhanced internal radiation flux of 100 and 1000 times the background at densities $\rm > 1~cm^{-3}$ due to the formation of stellar sources inside the halo. Figure \ref{figure4} shows the radial profiles for temperature, ionization degree of the gas, abundances of $\rm H_{2}$ and HD molecules. It is found that the temperature of the gas is relatively higher as compared to the cases with no internal radiation sources. The degree of ionization is also enhanced. The gas still cools and recombines inside the HII regions, which is why the degree of ionization declines towards the center of a halo. The $\rm H_{2}$ and HD abundances are shown in the bottom panel of figure \ref{figure4}. It is found that due to the enhanced strength of the internal UV flux, photodissociation of molecules becomes effective and $\rm H_{2}$ and HD are almost two orders of magnitude lower than in the no internal radiation case.
The state of the simulations for internal radiation sources at redshift 5.4 is shown in the figure \ref{figuredd}. It is found that the collapse of a halo is delayed for these cases. We find that for model E, gas is still able to cool down to a few hundred Kelvin, while for the lower dust-to-gas ratio case (i.e., model F) the temperature increases up to 1000 K. A binary system is again formed irrespective of the dust-to-gas ratio. The gas masses of disk 1 and 2 are $\rm 1.9 \times 10^{7}~M_{\odot}$ and $\rm 1.7 \times 10^{7}~M_{\odot}$ for $\rm Z/Z_{\odot}=10^{-2}$, and $\rm 1.6 \times 10^{7}~M_{\odot}$ and $\rm 8 \times 10^{6}~M_{\odot}$ for $\rm Z/Z_{\odot}=10^{-4}$. Despite the higher temperatures in the core of a halo due to photodissociation of molecules, surprisingly, fragmentation is enhanced for $\rm J_{21}=1000$ irrespective of the dust-to-gas ratio. This is due to cooling instabilities because the temperature lies in the unstable regime of 2000 to 5000 K. The formation and dissociation of $\rm H_{2}$ is very sensitive to the above-mentioned temperature range. As the temperature increases above 3700 K, the $\rm H_{2}$ survival time sharply decreases and the cooling time scale increases. On the other hand, if the temperature is slightly decreased, this leads to longer $\rm H_{2}$ survival times (see figure 4 of \cite{2002ApJ...569..558O}). This process initiates thermal instabilities. The gas masses of the fragments are on the order of a few $\rm 10^{6}~M_{\odot}$. Figure \ref{figured2} shows the temperature corresponding to the density projections in figure  \ref{figuredd} (i.e., models G and H). It can be seen that in the high density regions fine-grained thermal substructure occurs, which indicates small pockets of gas that quickly cool down to low temperatures. This supports our hypothesis of thermal instabilities leading to the local cooling and collapse of gas.

We also computed the ratio of rotational to circular velocity of the disks for internal radiation field cases (i.e., models E, F, G and H). These are shown in the top panels of figure \ref{figuret}. The ratio remains less than unity as in previous cases. We find that for a lower dust-to-gas ratio, the disks are now counter-rotating while they are co-rotating for higher dust-to-gas ratio case. Dynamical evolution is faster in high dust-to-gas ratio environments \citep{2011ApJ...737...63A}. Therefore, the lower-mass disk falls behind in its evolution compared to the higher-mass one for lower dust-to-gas ratio. Subsequently, the lower-mass disk is forced to accrete gas from the trailing spiral arm of the higher-mass disk and counter-rotation ensues. We also computed the turbulent and thermal energy of the system and found that turbulence is again highly supersonic and the disks are supported by turbulent pressure. The Toomre stability parameter Q is larger than unity in the central region (i.e., $\rm < 10^{19.6}~cm$), where we see a well-defined disk. The disk is therefore expected to be stable, similar to previous cases, as shown in the bottom panels of figure \ref{figuret}.

\subsection{Lyman Alpha emission from high-redshift protogalaxies}

\begin{figure*}[htb!]
\centering
\begin{tabular}{c c}
\begin{minipage}{8cm}
\includegraphics[scale=0.28]{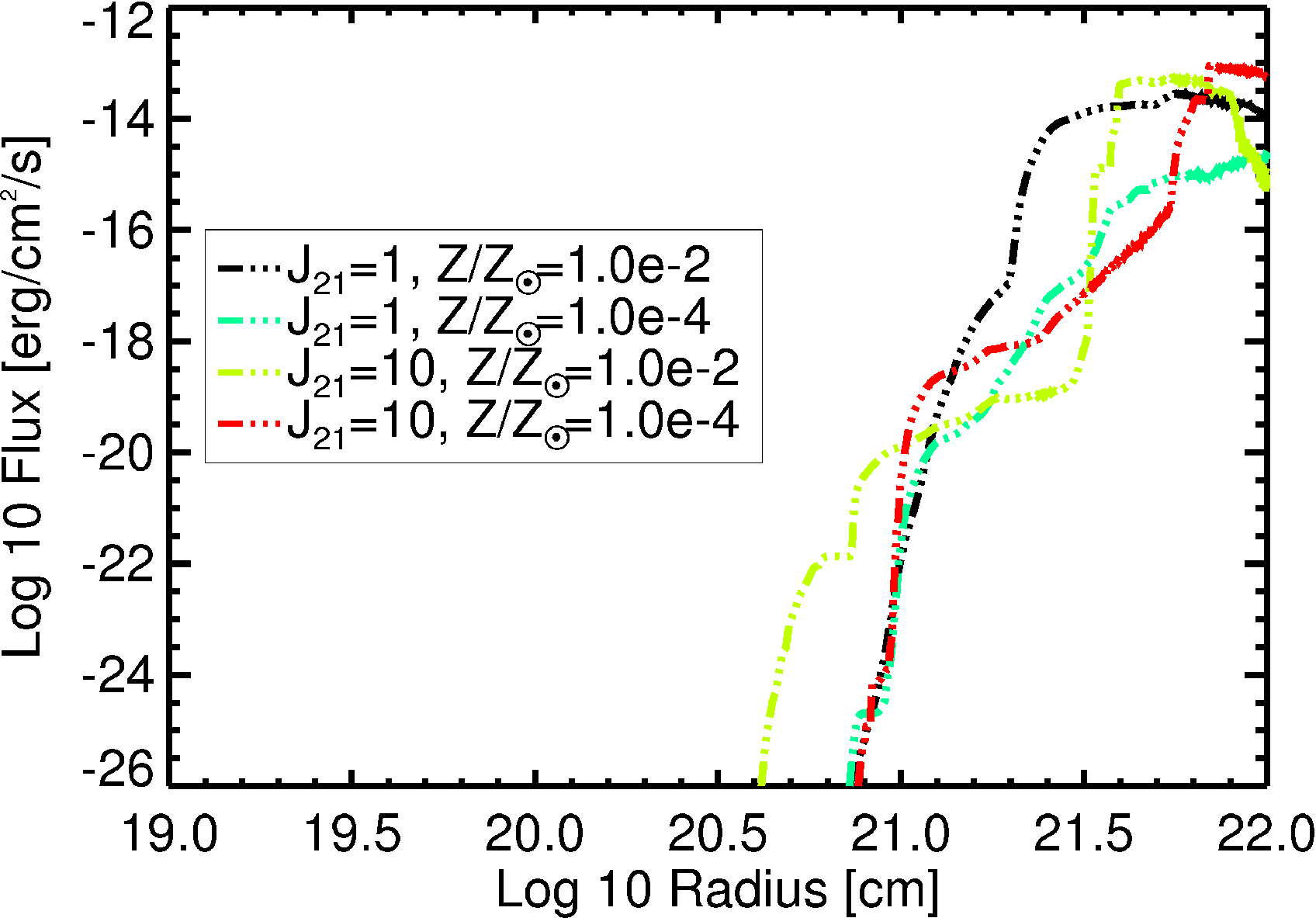}
\end{minipage} &
\begin{minipage}{8cm}
\includegraphics[scale=0.28]{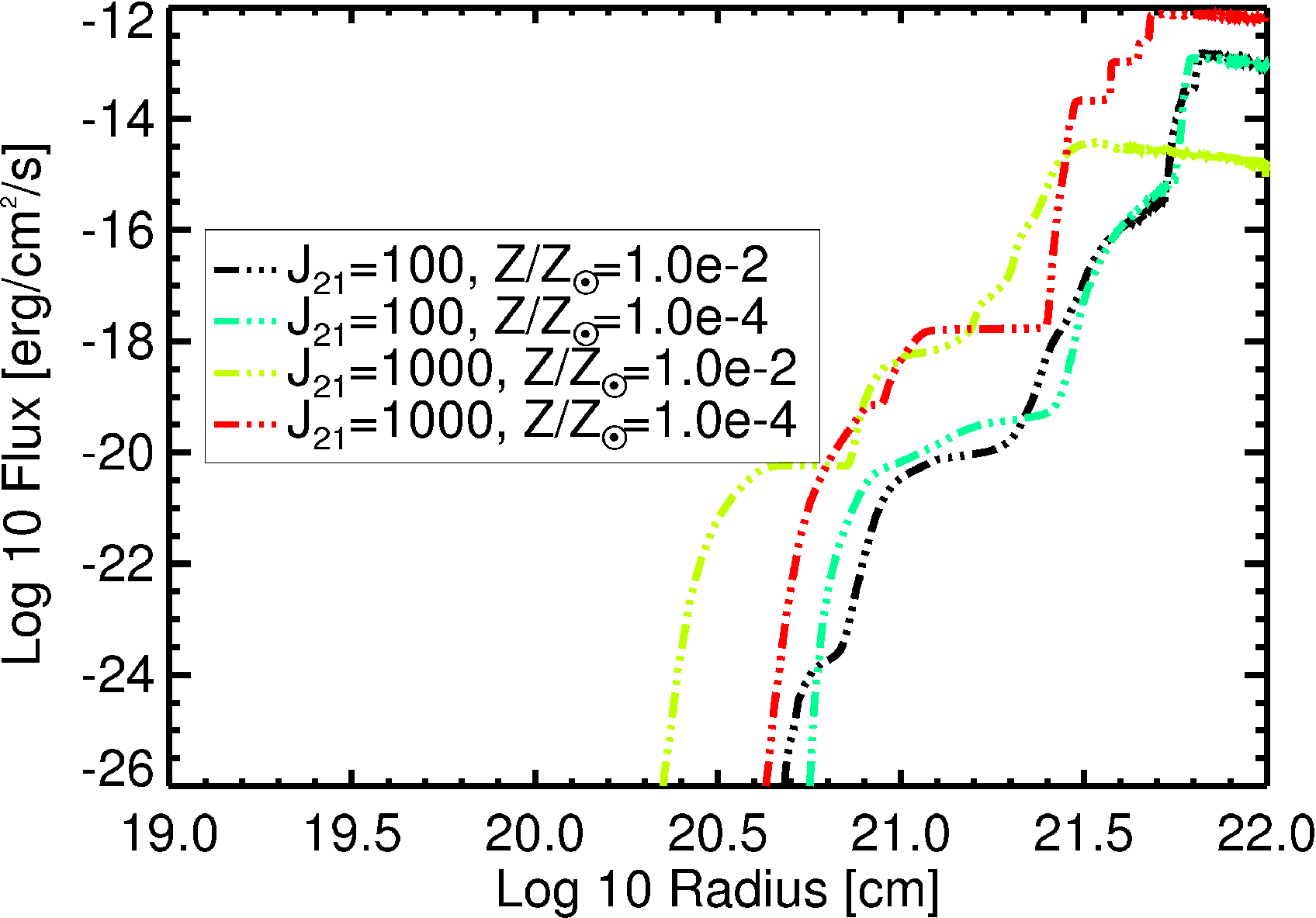}
\end{minipage} 
\end{tabular}
\caption{Lyman alpha emission for different magnitudes of the background UV field and dust-to-gas ratios. The right hand side panel shows the flux for different background UV radiation fields. The left panel shows the Lyman alpha flux for an internal UV radiation field in addition to a background UV flux $\rm J_{21}=1$. The value of the escape fraction for ionizing radiation is 10\% for both panels.}
\label{figuret2}
\end{figure*}

We estimate the Lyman alpha flux emanating from the galaxy forming halo. Our results for the Lyman alpha emission in the presence of dust are shown in figure \ref{figuret2}. We found that above columns of $\rm 10^{22}~cm^{2}$ and optical depths of $\rm \tau \geq 10^{7}$ trapping of Lyman alpha photons becomes effective, consistent with previous studies \citep{2006ApJ...652..902S,2011A&A...532A..66L,2011MNRAS.413L..33L,2011MNRAS.411.1659L}. That is why Lyman alpha emission sharply declines towards the center of the halo. It can be seen that the Lyman alpha flux emerges from the envelope of the halo. The value of the flux depends upon the strength of the external UV radiation field and the dust-to-gas ratio. The local variations in the flux are due to the clumpiness of the medium. We assume a constant dust-to-gas ratio, leading to a highly inhomogeneous, clumpy interstellar medium as a result of the high Mach numbers in our simulation \citep[see models of][]{1991ApJ...370L..85N,1999ApJ...518..138H,2011arXiv1102.1509S}. As dust sits in the center of the dense clumps, Lyman alpha photons scatter off these clumps and are able to escape the halo due to large scattering cross sections. This depends on the clump covering factor, i.e., average number of gas clouds along the line of sight. We compute the column density distribution and use it to determine the clump covering factor. We use the model of \cite{1999ApJ...518..138H} to compute the HI Lyman alpha escape fraction. Our findings are in agreement with \cite{1991ApJ...370L..85N,1999ApJ...518..138H,2011arXiv1102.1509S}.

The presence of a background UV flux excites the emission of Lyman alpha photons and enhances its luminosity. Moreover, flux predominantly arises from collisional excitation of atomic hydrogen and the contribution from recombination line emission is a few orders of magnitude lower. For further details on the origin and source of Lyman alpha emission see our previous studies \citep{2011MNRAS.413L..33L,2011A&A...532A..66L}. Our results also agree with previous studies that emission of Lyman alpha is extended and emerges from the warm gas in the surroundings of the galaxy \citep{2006ApJ...649...14D,2006ApJ...649...37D,2009MNRAS.400.1109D,2010ApJ...725..633F,2011A&A...532A..66L}. We found that a flux on the order of $\rm 10^{-14}-10^{-12}~erg/cm^{2}/s$ emanates from the the protogalactic halo at redshift 5.4. The redshifted wavelength of Lyman alpha photons at redshift 5.4 is 0.78 micron. This should be detectable with present-day telescopes like Subaru \footnote{http://www.naoj.org/Observing/Instruments/IRCS/camera/filters.html} and VLT \footnote{http://www.eso.org/public/teles-instr/vlt.html}. Further details on the issue of line trapping of Lyman alpha photons and the role of background UV flux in the emission of Lyman alpha photons are given by \cite{2010ApJ...712L..69S} and \cite{2011A&A...532A..66L}. Lyman alpha emission, apart from dust content, also depends on the velocity structure of a halo. The variation in Doppler velocity shift does not strongly influence the escape fraction of Lyman alpha photons \citep{1999ApJ...518..138H}. We also checked the consistency of intrinsic Lyman alpha luminosity with the adopted internal UV flux. We computed flux from intrinsic Lyman alpha luminosity by normalizing it with Lyman alpha line width and found it consistent with the adopted UV flux.
In the presence of dust, the contribution of the internal stellar feedback, on top of the background UV field, can vary strongly. For $Z/Z_\odot =10^{-4}$, UV continuum optical depths range from $\rm 10^{-6}$ to almost unity, and are a factor 100 times larger for $\rm Z/Z_{\odot} =10^{-2}$. In our simulations, column densities range from $\rm 10^{18}$ to $10^{24}$ $\rm cm^{-2}$, a similarly wide range as found by \cite{2009ApJ...702...63W}. For the dust-free case, we expect internal stellar feedback to be quite effective, and to partly photo-dissociate H$_2$. However, in the presence of a strong background UV field, the role of the intrinsic luminosity again diminishes.

\subsection{Dust emission from high-redshift protogalaxies}

We further compute the dust temperatures for different UV background fluxes and dust-to-gas ratios for the cases listed in table \ref{tab:sumary}. We found that for lower background UV fluxes (i.e., $\rm J_{21}=1$), the dust temperature remains close to the $\rm T_{CMB}$ while for higher fluxes (i.e., $\rm J_{21}=100$) it is well above $\rm T_{CMB}$. We have a maximum dust temperature of 34 K for the model H in our case. Our results are in agreement with observed dust temperatures of high-redshift quasars \citep{2001A&A...374..371O, 2011arXiv1110.1457Y}. We found that for the given dust-to-gas ratios and the halo masses considered here (dust masses, $\rm 10^6~M_{\odot}$ for  $\rm Z/Z_{\odot}=10^{-2}$ and $\rm 10^4~M_{\odot}$ for $\rm Z/Z_{\odot}=10^{-4}$), the dust continuum emission is far below the detection limits of the Atacama Large Millimeter Array (ALMA)\footnote{http://www.almaobservatory.org/en/science-with-alma/evolved-stars-stardust} at z=5.4. However, ALMA will be able to explore the dust content in larger systems or in the presence of higher dust-to-gas ratios. Our calculations show that a 3 sigma detection is possible with ALMA band 9 in five hours of integration if dust masses (for calculation see equation 1 of \cite{2001A&A...374..371O}) of $\rm 6.75 \times 10^{8}~M_{\odot}$, $\rm 3.27 \times 10^{8}~M_{\odot}$, and $\rm 1.54 \times 10^{8}~M_{\odot}$ are present (assuming dust temperatures of 20 K, 30 K, and 50 K, respectively). One requires roughly solar metallicity for a $\rm \sim 10^{10}~M_{\odot}$ protogalactic halo to be detectable at $\rm z>5$.
% In order to observe the smaller, non-QSOs, galaxies that we model, one needs to have higher dust abundances and higher dust temperatures.  Dust masses (\ch{for calculation see equation 1 of \cite{2001A&A...374..371O}}) that can be observed with ALMA band 9, for an integration time of 5 hours at a flux of 0.1 mJy (3 sigma detection) and redshift 5.4, are, for dust temperatures of 20 K, 30 K and 50 K, $\rm 6.75 \times 10^{8}~M_{\odot}$, $\rm 3.27 \times 10^{8}~M_{\odot}$ and $\rm 1.54 \times 10^{8}~M_{\odot}$, respectively.

\section{Discussion and Conclusions}

In all, we have performed 9 cosmological simulations to study the dynamics of disk formation in protogalactic halos under the influence of different UV radiation field strengths and dust-to-gas ratios. We used the adaptive mesh refinement code FLASH to carry out this task and coupled it with a chemical network. Our chemical model includes photoionization, collisional excitation, collisional ionization, radiative recombination, photodissociation processes and the formation of molecules on dust grains as well as a multi-level treatment of the hydrogen atom.
We found that gas in the atomic cooling halo is photoionized by a background UV flux and that photoheating raises the gas temperature to above $\rm 10^{4}$ K. At densities $\rm \geq 10^{-24}~g/cm^{-3}$, gas cools, recombines and falls into the center of a halo through cold streams of gas. The later thermal evolution of a halo depends on the strength of a background UV flux and the dust-to-gas ratio. The formation of $\rm H_{2}$ and HD molecules is significantly enhanced (by two orders of magnitude) in the presence of dust grains as compared to gas phase reactions. We found that the formation of molecules is more effective for higher dust-to gas ratio (i.e., $\rm Z/Z_{\odot}=10^{-2}$) as compared to lower dust-to-gas ratio (i.e., $\rm Z/Z_{\odot}=10^{-4}$). The minimum temperature in the core of a halo is $\rm \sim$ 200 K for model A. This value is higher (800 K) for lower dust-to-gas ratio and stronger UV radiation flux. We conclude that the thermal evolution of a protogalactic halo is strongly influenced by the presence of dust and a UV radiation field.

Our results show that binary disks are formed in the protogalactic halo at redshift 5.4 irrespective of the background UV radiation field strength and dust-to-gas ratio in our simulations. However, the morphology and the formation time of disks does strongly depend on the dust-to-gas ratio and strength of a background UV flux. We found that disks are stable as the value of Toomre instability parameter Q is greater than one at radii $\rm <10^{19.5}$ cm. We also found that the turbulence in the halo is highly supersonic due to the higher virial temperature of the halo. Moreover, disks are more supported by turbulent pressure than thermal pressure. It may happen that disks merge with one another at lower redshifts. There is also a possibility that more than 2 disks are formed in some cases, which will lead to the formation of a multiple disk system. We could not follow the evolution of a halo to a lower redshift than 5.4 as this becomes a computationally too expensive task. Nevertheless, we have explored an important phase of disk formation, which may be observed with future planned surveys.

We also performed a set of simulations with different Gaussian random field initial conditions to determine whether the origin of binary disk formation comes from cosmological initial conditions. We found that a single disk is formed in the center of a halo instead of binary disks for a different random seed. We thus think that the origin of binary disk formation is associated with the underlying dynamics of the dark matter in the halo. In an upcoming paper, we will address the issue of how often these binary disks form in atomic cooling halos.

We performed additional simulations for the cases where the background UV flux is enhanced by factors of 100 and 1000 due to the formation of radiation sources like stars inside the halo. It was found that the collapse of a halo is delayed in HII regions formed by the internal radiation sources and the photodissociation of molecules ($\rm H_{2}$ and HD) is strongly enhanced in this case. We found that despite the higher temperatures in the halo for an internal radiation flux of $\rm J_{21}=100$, in addition to a background UV flux of $\rm J_{21}=1$, again binary disks are formed for both $\rm Z/Z_{\odot}=10^{-2}$ and $\rm Z/Z_{\odot}=10^{-4}$. For a higher strength of the internal radiation flux, $\rm J_{21}=1000$, the temperature of the halo in the center is increased to a few thousand Kelvin. We found that for this case cooling instabilities lead to the formation of multiple clumps for both dust-to-gas ratios, $\rm Z/Z_{\odot}=10^{-2}$ and $\rm Z/Z_{\odot}=10^{-4}$ at redshift 5.4. Due to significant local overdensities in the high-Mach turbulence, we expect a star formation mode that is regulated by the turbulent density PDF \citep{2004RvMP...76..125M,2007ApJ...654..304K,2012MNRAS.tmp.2590B}. The further fate of these clumps depends on how they transfer angular momentum. They may merge with each other and form a disk if they lose angular momentum. It may also happen that they keep cooling, and collapsing, and fragment further.

We estimated the Lyman alpha emission from protogalactic halos for different strengths of UV radiation fields and dust-to-gas ratios. We found that even in the presence of dust, for $\rm Z/Z_{\odot}=10^{-2}$ and $\rm Z/Z_{\odot}=10^{-4}$, Lyman alpha photons scatter off dusty gas clouds and an inhomogeneous dust distribution leads to the efficient escape of Lyman alpha photons. Our results are in agreement with the models of \cite{1999ApJ...518..138H}. It was found that a flux of the order of $\rm 10^{-14}-10^{-12}~erg/cm^{2}/s$ emerges from the envelope of the halo due to effective line trapping of Lyman alpha photons in the center of a halo, consistent with previous studies \citep{2006ApJ...652..902S,2010ApJ...712L..69S,2011A&A...532A..66L,2011MNRAS.413L..33L}. Such a flux should be observable with present-day telescopes like Subaru and the VLT. Our results provide a conservative lower limit to the Lyman alpha flux as we have computed the flux for minimum velocity gradient along the line of sight. Radiative transfer of Lyman alpha photons strongly depends on the amount of structure in the halo as well as on velocity gradients. Therefore, cosmological simulations coupled with Lyman alpha radiative transfer should be performed in the future to obtain more robust results.

Finally, we computed the dust emission from the protogalactic halo in our simulations. We found that due to the low dust masses such systems cannot be observed with ALMA at z=5.4. Only galaxies with higher metal content and high dust temperatures can be probed. However, we also found that dust masses of few $\rm \times 10^{8}~M_{\odot}$ yield $\rm \sim 0.1$ mJy of flux and thus a 3-sigma detection in 5 hours. Given that our protogalactic halos are likely to build up a solar content of metals within 1-2 Gyr, they should be detectable for $\rm z\leq 5$.

The formation of disks may also have significant implications for magnetic fields in the first galaxies. In the absence of rotation, the small-scale dynamo is expected to produce tangled magnetic fields, close to equipartition with turbulent energy \citep{2009A&A...494...21A,2010A&A...522A.115S}. This process was shown to be efficient for a large range of turbulence models and Mach numbers \citep{2011PhRvL.107k4504F,2012PhRvE..85b6303S}. The presence of rotating disks, on the other hand, gives rise to large-scale dynamo effects amplifying the mean magnetic field \citep{2005PhR...417....1B}. The presence of magnetic fields in high-redshift galaxies will be probed in the future with the SKA\footnote{http://www.skatelescope.org}

\section*{Acknowledgments}

The FLASH code was in part developed by the DOE-supported Alliance Center for Astrophysical Thermonuclear Flashes (ACS) at the University of Chicago. We thank Stephanie Cazaux, Seyit Hocuk, and Gustavo Dopcke for valuable discussions on the topic. DRGS acknowledges funding from the DFG priority program 1573 (project number SCHL 1964/1-1) and the SFB~963 (project A12). DRGS also thanks the German Science Foundation for funding via the SPP 1573 (project number SCHL~1964/1-1) and the SFB 963/1 {\em Astrophysical Flow Instabilities and Turbulence}. We also thank the anonymous referee for a careful reading of the manuscript and useful feedbacak. Our simulations were carried out on the Gemini machines at the Kapteyn Astronomical Institute, University of Groningen.

\label{lastpage}

\bibliography{biblio5.bib}

\end{document}